\documentclass[useAMS,usenatbib]{mn2e}

\usepackage{graphicx}	% Including figure files
\usepackage{amsmath}	% Advanced maths commands
\usepackage{amssymb}	% Extra maths symbols
\usepackage{multicol}        % Multi-column entries in tables
\usepackage{bm}		% Bold maths symbols, including upright Greek
\usepackage{pdflscape}	% Landscape pages
\usepackage{multirow,multicol} 

\def\N1316{NGC\,1316}
\def\N1404{NGC\,1404}

\def\4U{4U~1735$-$444}

\def\arcsec{\ifmmode '' \else $''$\fi}

\def\arcsecpoint{\ifmmode ''\!. \else $''\!.$\fi}

\def\kms{\ifmmode {\rm km\ s}^{-1} \else km s$^{-1}$\fi}
\def\Msun{\ifmmode {\rm M}_{\odot} \else M$_{\odot}$\fi}
\def\Lsun{\ifmmode {\rm L}_{\odot} \else L$_{\odot}$\fi}
\def\Zsun{\ifmmode {\rm Z}_{\odot} \else Z$_{\odot}$\fi}

\def\ergscm2{ergs\,s$^{-1}$\,cm$^{-2}$}
\def\icm3{{\rm cm}^{-3}}
\def\icm2{{\rm cm}^{-2}}
\def\qo{\ifmmode q_{\rm o} \else $q_{\rm o}$\fi}
\def\Ho{\ifmmode H_{\rm o} \else $H_{\rm o}$\fi}
\def\ho{\ifmmode h_{\rm o} \else $h_{\rm o}$\fi}

\def\vFWHM{\ifmmode v_{\mbox{\tiny FWHM}} \else
            $v_{\mbox{\tiny FWHM}}$\fi}
\def\CCF{\ifmmode F_{\it CCF} \else $F_{\it CCF}$\fi}
\def\ACF{\ifmmode F_{\it ACF} \else $F_{\it ACF}$\fi}
\def\Halpha{\ifmmode {\rm H}\alpha \else H$\alpha$\fi}
\def\Hbeta{\ifmmode {\rm H}\beta \else H$\beta$\fi}
\def\Hgamma{\ifmmode {\rm H}\gamma \else H$\gamma$\fi}
\def\Hdelta{\ifmmode {\rm H}\delta \else H$\delta$\fi}
\def\Lya{\ifmmode {\rm Ly}\alpha \else Ly$\alpha$\fi}
\def\Lyb{\ifmmode {\rm Ly}\beta \else Ly$\beta$\fi}
\def\Lyg{\ifmmode {\rm Ly}\beta \else Ly$\gamma$\fi}

\def\ciii{\ifmmode {\rm C}\,{\sc iii} \else C\,{\sc iii}\fi}
\def\civ{\ifmmode {\rm C}\,{\sc iv} \else C\,{\sc iv}\fi}
\def\cv{\ifmmode {\rm C}\,{\sc v} \else C\,{\sc v}\fi}
\def\cvi{\ifmmode {\rm C}\,{\sc vi} \else C\,{\sc vi}\fi}

\def\nvii{N\,{\sc vii}}

\def\o5007{[O\,{\sc iii}]\,$\lambda5007$}

\def\ovii{O\,{\sc vii}}
\def\oviii{O\,{\sc viii}}
\def\oviiviii{O\,{\sc vii-viii}}

\def\neix{Ne\,{\sc ix}}
\def\nex{Ne\,{\sc x}}
\def\neixx{Ne\,{\sc ix-x}}

\def\mgxi{Mg\,{\sc xi}}

\def\sixiii{Si\,{\sc xiii}}

\def\fexvii{Fe\,{\sc xvii}}

\def\fexviii{Fe\,{\sc xviii}}
\def\fexxii-iii{Fe\,{\sc xxii-xxiii}}

\usepackage{xcolor}

\usepackage{hyperref}

\hypersetup{colorlinks=true,linkcolor=blue,citecolor=blue,filecolor=blue,urlcolor=blue}

\title[The outflows of NGC 247 ULX-1]
{\centering{XMM-\textit{Newton} campaign on the ultraluminous X-ray source NGC 247 ULX-1: outflows}}
\author[C. Pinto et al.]{C. Pinto,$^{1,2}$\thanks{E-mail:
ciro.pinto@inaf.it} R. Soria,$^{3}$ D. J. Walton,$^{4}$ A. D'A\`{i},$^{1}$ F. Pintore,$^{1}$ P. Kosec,$^{5}$ \newauthor
W. N. Alston,$^{6}$ F. Fuerst,$^{6}$ M. J. Middleton,$^{7}$ T. P. Roberts,$^{8}$ M. Del Santo,$^{1}$ \newauthor
D. Barret,$^{9}$ E. Ambrosi,$^{1}$ A. Robba,$^{1}$ H. Earnshaw,$^{10}$ and A. C. Fabian$^{4}$\\
$^{1}$INAF - IASF Palermo, Via U. La Malfa 153, I-90146 Palermo, Italy\\
$^{2}$ESTEC/ESA, Keplerlaan 1, 2201AZ Noordwijk, The Netherlands\\
$^{3}$College of Astronomy and Space Sciences, University of the Chinese Academy of Sciences, Beijing 100049, China\\
$^{4}$Institute of Astronomy, Madingley Road, CB3 0HA Cambridge, United Kingdom\\
$^{5}$MIT Kavli Institute for Astrophysics and Space Research, Cambridge, MA 02139, USA\\
$^{6}$ESAC/ESA European Space Astronomy Center, P.O. Box 78, 28691 Villanueva de la Canada, Madrid, Spain\\
$^{7}$Physics \& Astronomy, University of Southampton, Southampton, Hampshire SO17 1BJ, UK\\
$^{8}$Centre for Extragalactic Astronomy, Durham University, Department of Physics, South Road, Durham DH1 3LE, UK\\
$^{9}$Universit\'e de Toulouse, CNRS, IRAP, 9 Avenue du colonel Roche, BP 44346, 31028 Toulouse Cedex 4, France\\
$^{10}$Cahill Center for Astronomy and Astrophysics, California Institute of Technology, Pasadena, CA 91125, USA}

\begin{document}

\date{Accepted 2021 June 3. Received 2021 May 28; in original form 2021 April 22.}

\pagerange{\pageref{firstpage}--\pageref{lastpage}} \pubyear{2018}

\maketitle

\label{firstpage}

\begin{abstract}
Most ULXs are believed to be powered by super-Eddington accreting neutron stars 
and, perhaps, black holes. Above the Eddington rate the disc is expected to thicken
and to launch powerful winds through radiation pressure.
Winds have been recently discovered in several ULXs. 
However, it is yet unclear whether the thickening of the disc or the wind variability causes
the switch between the classical soft and supersoft states observed in some ULXs.
In order to understand such phenomenology and the overall super-Eddington 
mechanism, we undertook a large (800\,ks) observing campaign with XMM-\textit{Newton} to study 
NGC 247 ULX-1, which shifts between a supersoft and classical soft ULX state.
The new observations show unambiguous evidence of a wind in the form
of emission and absorption lines from highly-ionised ionic species, with the latter
indicating a mildly-relativistic outflow ($-0.17c$) in line with the detections in other ULXs. 
Strong dipping activity is observed in the lightcurve and primarily during the 
brightest observations, which is typical among soft ULXs, and indicates
a close relationship between the accretion rate and the appearance of the dips.
The latter is likely due to a thickening of the disc scale-height and the wind as shown by a 
progressively increasing blueshift in the spectral lines. 
%%%Forthcoming papers will
%%%This work was the first in a series and more papers using focussing on marvellous dataset
%%%are being prepared to obtain further insights on the broadband spectral evolution, the 
%%%corresponding variability timescales, and the source phenomenology.
\end{abstract}


\section{Introduction}
\label{sec:intro}

There is a consensus that the majority of ultraluminous X-ray sources (ULXs) are stellar-mass compact objects (neutron stars and perhaps black holes) accreting above the critical Eddington rate (see, {\it e.g.}, \citealt{King2001}, \citealt{Poutanen2007}; \citealt{Middleton2011}, \citealt{Bachetti2014}, \citealt{Kaaret2017}). The spectral curvature in the X-ray band, with a characteristic downturn above $\sim$5 keV, and the presence of residuals at energies $\lesssim$1 keV \textcolor{black}{when modelled with featureless continuum models}, are two characteristic ULX features (see, {\it e.g.}, \citealt{Soria2004}, \citealt{Goad2006}, \citealt{Gladstone2009}) that are naturally explained if the primary X-ray photons are seen through a disc wind, as expected for systems accreting in the super-Eddington regime ({\it e.g.}, \citealt{SS1973}, \citealt{Poutanen2007}). The thickness of the wind and, as a result, the ``softness'' of the observed ULX spectra in the $\sim$0.1--20 keV band, are likely a function of two main parameters: the mass outflow rate in the wind (which is related to the accretion rate), and our viewing angle, with softer sources being observed closer to the disc plane (see, {\it e.g.}, \citealt{Sutton2013}, \citealt{Middleton2015a}, \citealt{Pinto2017}).  In particular, ULXs with a power-law photon index $\Gamma > 2$ in the $\sim$0.3--5 keV band are empirically classified in a ``soft ultraluminous'' (SUL) state.

\subsection{Supersoft ultraluminous sources}

Among the ULXs, a special sub-class is represented by ultraluminous supersoft sources (ULSs), also referred to as sources in the ``supersoft ultraluminous'' (SSUL) state \citep{Feng2016}. This state is defined by a dominant, cool blackbody component ($T_{\rm bb} \lesssim 140$ eV), with very weak or completely absent hard component at higher energies. Observationally, sources in the SSUL state show essentially no photons $>$1 keV. 
Their bolometric luminosity is typically a few $10^{39}$ erg s$^{-1}$. They should not be confused with the ``classical" supersoft sources (see, {\it e.g.}, \citealt{Krautter1996}), which are usually interpreted as nuclear burning on the surface of a white dwarf.  Classical supersoft sources also have blackbody spectra, but at lower luminosities ($L_{\rm bb} \lesssim 10^{38}$ erg s$^{-1}$), and generally with a smaller bb radius: $R_{\rm bb} \sim 5000-10000$ km, consistent with a white dwarf, while SSUL spectra can have blackbody radii as large as $10^5$ km (\citealt{Urquhart2016}).  

Blackbody modelling of SSUL spectra at different epochs shows that the characteristic radius is not constant, which rules out a hard surface as the origin of the thermal emission \citep{Feng2016}. It also shows a clear anti-correlation between blackbody radius and temperature, which rules out a standard accretion disc. Instead, such behaviour is consistent with emission from the photosphere of an optically-thick wind. The increase in the photospheric radius corresponds to an enhancement of the wind thickness. 
%%%The high-temperature limit of the SSUL regime appears to be at $T_{\rm bb} \approx 140$ eV, with a characteristic radius of a few 1000 km (\citealt{Urquhart2016}).
%%%The low-temperature limit is not well constrained, because when $T_{\rm bb} \lesssim 40$ eV (associated with a radius of $\sim10^5$ km), the source is no longer detectable with either \textit{Chandra} or XMM-\textit{Newton} (or any other current space telescope). For the same reason, we do not yet know which of those sources are genuinely transient, and which simply disappear from our view at times when their photosphere becomes too large and cool.

The apparent non-periodic spectral variability in the SSUL state (\citealt{Liu2008}) is very likely due to a thickening of the wind along our line of sight. One question we want to address is whether the short-term variations in optical depth are just stochastic variability (at a given mass accretion rate $\dot{M}$) due to the clumpy nature of the wind (``weather", see, {\it e.g.}, \citealt{Takeuchi2013}), or instead each variation is driven by a change in the underlying $\dot{M}$ (``climate"), which then affects the wind density and launching radius. 
%%%This requires to study the relationship between the column density of the wind and the luminosity of the source, which we obtain from flux-resolved spectroscopy and a comparison among different sources (\citealt{Feng2016}; \citealt{Urquhart2016}).

%%%In fact, the bolometric luminosity of several SSUL sources slightly increases 
%%%when the blackbody temperature decreases and the blackbody radius increases 
%%%This is a hint that increases in the photospheric radius correlate with temporary enhancements of the accretion rate in the wind-launching region.
%%%, but it is not easy to disentangle this forcing from short-term stochastic variability. 

%%%To advance our understanding of the SSUL regime, we need to look at those few spectral features that differ from a pure blackbody such as the high-energy ($>1$ keV) tail, the strong absorption feature just above 1 keV.
%%%%%%, significant even with the low spectral resolution of CCD detectors.
%%%A power-law tail is detected in many sources (see, {\it e.g.}, \citealt{Kong2006}). Crucially, the tail appears or is stronger when the thermal component is more compact and hotter. It disappears or fades when the thermal component becomes larger/cooler. This suggests that the bloating of the photosphere is associated with a stronger, thicker wind, and the complete obscuration of the hard photons. 

A deep absorption edge was detected at $\approx$1.0--1.1 keV in several SSUL sources ({\it {\it e.g.}}, those in M\,51, NGC\,6946 and M\,101: \citealt{Urquhart2016}, \citealt{Earnshaw2017}). This feature clearly appears when the source softens, progressively losing all X-ray photons with energy above 1 keV. There are no strong absorption edges predicted at the observed energy of those dips. Viable solutions are blueshifted {\oviii} ionisation edges (871eV at rest), which was observed in novae during their supersoft phase ({\it e.g.}, \citealt{Pinto2012b}), or a combination of high-ionisation Fe\,L or Ne\,{\sc ix-x} absorption lines with velocities of $\sim$0.1--0.2c, supporting the case for an optically-thick, mildly-relativistic wind.
 
\subsection[]{NGC 247 ULX-1}

%%%For the next crucial step of our study, we identified and studied a ULX that switched between the SUL and the SSUL state: NGC\,247 ULX-1. This source gives us the unique chance to determine how the power-law tail and the line features appear and disappear during those transitions. 

Previous X-ray studies (see, {\it e.g.}, \citealt{Feng2016}), based on two short XMM-\textit{Newton} observations, showed that NGC\,247 ULX-1 switched from a supersoft state with hardly any flux above 1 keV (in 2009) to a much brighter state (in 2014) consistent with the soft end of the ``standard'' ULX population ({\it i.e.}, those with a bright hard X-ray spectral tail). 
%%%For example, in its brightest state, NGC\,247 ULX-1 appears very similar to the previously well studied soft ULX in NGC\,55, but with a 2-3 times higher luminosity, peaking at $L_{\rm X} \approx 7 \times 10^{39}$ erg s$^{-1}$ (\citealt{Pintore2015}, \citealt{Pinto2017}). This suggests that the SSUL sources are a sub-type of ULXs with extreme wind properties, rather than a separate physical class of accreting binaries.

NGC\,247 ULX-1 is also among the most variable ULXs. It exhibits strong dips in its X-ray lightcurve, which last several ks, and during which the flux decreases by an order of magnitude \citep{Feng2016}. Both the low and high flux spectra are characterised by strong \textcolor{black}{spectral features}. %%%, including the absorption edge at $\approx$1.0 keV.

\citet{Pinto2017} found similarities between the narrow X-ray spectral features of NGC\,247 ULX-1 in the high flux state with those of NGC\,55 ULX, using observations taken with the high-resolution gratings aboard XMM-\textit{Newton}. 
Despite the short duration (33 ks) of the observations available, two remarkable absorption features at 7.5\,{\AA} and 16.2\,{\AA} were found along with other weaker features, which can be modelled with absorption from photoionised gas outflowing at $\approx0.14c$. The observation taken in the low-flux (supersoft) state is far too short to provide any useful data. Thus, we proposed and were awarded a 800\,ks deep XMM-\textit{Newton} programme to characterise the properties of the outflows and the origin of the spectral residuals, and correlate them with the continuum flux variability and state changes.

This paper is the first in a series of intriguing results from our XMM-\textit{Newton} campaign. Here, we focus on the search for wind signatures in the time-average spectrum, taking full advantage of the high-spectral-resolution data, and on the variability of the spectral features around 1 keV. We detail our observing campaign in Sect. \ref{sec:xmm_campaign} and present the results of our spectral analysis in Sect. \ref{sec:spectral_analysis}. We discuss the results in Sect. \ref{sec:discussion}, and outline some conclusions in Sect. \ref{sec:conclusion}.
%%%Technical details are reported in Appendix \ref{sec:appendix}. 

\section[]{NGC 247 XMM-\textit{Newton} campaign}
\label{sec:xmm_campaign}

We observed the NGC 247 galaxy between December 2019 and January 2020.
The roll angle was similar throughout the whole campaign and avoided strong contamination 
along the dispersed grating spectra from the nearby brightest X-ray sources 
(see Fig.\,\ref{Fig:Plot_EPIC_image} and Appendix \ref{sec:appendix_source_x2} for more detail).
Seven observations were expected to occur but, owing to an issue with the RGS instrument that
occurred during the last observation (id:0844860701), an additional final
observation was taken shortly afterwards (id:0844860801) to recover the lost exposure.
In Table\,\ref{table:obs_log}, we report the detail of our observations.
We performed the spectral analysis with the {\scriptsize{SPEX}}
code \citep{kaastraspex}, we used C-statistics (C-stat, \citealt{Cash1979}) for spectral fits, 
which was proved to be efficient in comparing models similarly to $\chi^2$ statistics
(\citealt{Kaastra2017}), and we adopted 1-$\sigma$ confidence intervals. 

\begin{table}
\caption{XMM-\textit{Newton} campaign on NGC 247 ULX-1. }  
\label{table:obs_log}     
\renewcommand{\arraystretch}{1.1}
 \small\addtolength{\tabcolsep}{-3pt}
 \vspace{-0.1cm}
\scalebox{0.95}{%
\hskip-0.2cm\begin{tabular}{@{}ccccccc}     
\hline  
OBS\_ID  &  Date   & t$_{\rm RGS 1}$ & t$_{\rm RGS 2}$ & t$_{\rm MOS1}$ & t$_{\rm MOS2}$ & t$_{\rm pn}$  \\
                &             & (ks)                     & (ks)                     & (ks)                     & (ks)                     & (ks)                \\
\hline                                                                                                  
 0844860101 & 2019-12-03 & 110.5 & 110.1 & 104.6 & 105.0 & 76.6 \\
 0844860201 & 2019-12-09 & 110.9 & 110.6 & 109.3 & 109.2 & 90.3 \\
 0844860301 & 2019-12-31 & 117.4 & 117.0 & 112.6 & 113.9 & 76.8 \\
 0844860401 & 2020-01-02 & 112.3 & 112.0 & 110.6 & 110.5 & 93.5 \\
 0844860501 & 2020-01-04 & 115.9 & 115.6 & 113.3 & 113.3 & 94.4 \\
 0844860601 & 2020-01-06 & 102.3 & 101.7 &  83.3 &  83.2 & 57.4 \\
 0844860701 & 2020-01-08 &  28.2 &  28.3 &  96.1 &  97.5 & 70.1 \\
 0844860801 & 2020-01-12 &  61.0 &  60.8 &  59.3 &  59.4 & 41.7 \\
\hline                
 Total  [ks]          &                 & 758.5 & 756.1 & 789.1  & 792.0 & 600.8 \\
 Total  [kcnts] &                 &     7.2 &   10.4 &   56.7  &   57.2 & 186.5 \\
\hline                
\end{tabular}}

Notes: exposure times account for high background removal. 
Source counts are in the whole energy band for each detector.
\vspace{-0.3cm}
\end{table}

\subsection[]{Data preparation}
\label{sec:data_reduction}

In this work we used data from the European Photon-Imaging Camera 
(EPIC), the Reflection Grating Spectrometer (RGS) and the Optical Monitor (OM) 
aboard XMM-\textit{Newton}.
The primary science was carried out 
with the RGS which can detect and resolve narrow 
spectral features. The broadband cameras (EPIC MOS 1,2 and pn) were mainly
used to determine the spectral continuum and cover the hard X-ray band
missed by the RGS. 
%%%The EPIC count rate is much higher than RGS
%%%in the soft ($<\,2$ keV) band, but the EPIC energy resolution 
%%%is insufficient to resolve narrow features, particularly in the soft band.
%%%($R_{\,0.3-2 \, \rm keV, \, EPIC}=\Delta E/E \sim 10-20$). 
%Therefore, our main wind search consists of using EPIC spectra between 
%2 and 10 keV, and RGS data below 2 keV
%($R_{\,0.3-2 \, \rm keV, \, RGS}=\Delta E/E \sim 150-600$).

\subsubsection{EPIC cameras}
\label{sec:data_reduction_epic}

We reduced the 8 new XMM-\textit{Newton} observations with the 
Science Analysis System ({\scriptsize{SAS}} v18.0.0,
%\footnote{https://www.cosmos.esa.int/web/xmm-newton/sas}
CALDB available on March, 2020).
EPIC-pn and MOS data were reduced with the {\scriptsize{EPPROC}} and
{\scriptsize{EMPROC}} tasks, respectively.
Following the recommened procedures,
we filtered the MOS and pn event lists for bad pixels, cosmic-ray events 
outside the field of view, photons in the gaps (FLAG$=$0), 
and applied standard grade selections (PATTERN $\leq12$
for MOS and PATTERN $\leq4$ for pn).
We corrected for contamination from high background 
by selecting background-quiescent intervals
on the lightcurves for MOS\,1,2 and pn in the 10$-$12 keV
energy band. 
\textcolor{black}{These} lightcurves were extracted in time bins of 100\,s and all those
with a count rate above 0.4 c/s for pn and 0.2 c/s for MOS were rejected. 
The MOS\,1-2 and pn clean exposure times are 
reported in Table\,\ref{table:obs_log}.

We extracted EPIC MOS 1-2 and pn images in the 0.3$-$10\,keV energy range
and stacked them with the {\scriptsize{EMOSAIC}} task
(see Fig.\,\ref{Fig:Plot_EPIC_image}).
We also extracted EPIC MOS 1-2 and pn lightcurves
from within a circular region of 20'' 
radius centred on the source position ($\alpha,\delta$=00:47:04.0,-20:47:45.7).
We used the task {\scriptsize{EPICLCCORR}}, which corrects for
vignetting, bad pixels, chip gaps, PSF, and 
quantum efficiency.
The background (BKG) lightcurves were extracted from within a larger circle 
in a nearby region on the same chip.
%%%, but away from bright sources and the readout direction.
%%%We also made sure that the background region was not placed in
%%%the copper emission region (\citealt{Lumb2002}).
In order to measure the variations in the spectra hardness, we extracted EPIC-pn 
(for its larger effective area) lightcurves
in the soft (0.3--1 keV) and hard (1--10 keV) energy bands (with 1 ks time bins to increase
the signal-to-noise ratio of each bin).
\textcolor{black}{The PN lightcurves are shown in Fig.\,\ref{Fig:Plot_XMM_lc} (top-left panel)
with a length above 800 ks (despite an effective clean exposure of 600ks) due to the large 1ks bin size
which does not show the time gaps of high BKG lasting few 100s.
The lightcurves were also glued together for displaying purposes but with vertical dotted lines
separating them.}
The hardness ratio was defined as the fraction of hard photons with respect to the total
($H/(H+S)$). The boundary energy was adopted owing to the strong spectral curvature
of supersoft sources exhibited above 1 keV (see Sect. \ref{sec:data_investigation}).
%%%These lightcurves are shown in Fig.\,\ref{Fig:Plot_XMM_lc}.

We extracted EPIC MOS 1-2 and pn spectra in the same regions used 
for the lightcurves. 
%%%We used the task {\scriptsize{EVSELECT}} for the extraction
%%%and the tasks {\scriptsize{ARFGEN}} and {\scriptsize{RMFGEN}} to compute the effective area
%%%and the response matrix for each individual observation. 
The EPIC-pn spectra of the individual observations 
are shown in Fig.\,\ref{Fig:Plot_EPIC_spectra}. We avoided over-plotting the MOS
spectra for displaying purposes.

\subsubsection{RGS cameras}
\label{Sect:rgs_reduction}

The RGS data reduction was performed with the {\scriptsize{RGSPROC}}
pipeline. We filtered out periods affected by contamination from Solar flares 
by selecting background-quiescent intervals in the lightcurves of the RGS 1,2 CCD\,9
({\it i.e.}, $\gtrsim1.7$\,keV) with a count rate below 0.2 c/s. 
As usual, Solar flares affected the RGS data on a much lower level than EPIC.
The total clean exposure times are 
quoted in Table\,\ref{table:obs_log}.

We extracted the $1^{\rm st}$ and $2^{\rm nd}$-order RGS spectra in a cross-dispersion region 
of 0.8' width, centred on the source coordinates and the background spectra by selecting photons
beyond the 98\% of the source point-spread-function.
The background regions do not overlap with bright sources. 
%%%We also checked for consistency with the background produced 
%%%using the template background model from blank field data
%%%(obtained from the {\scriptsize{RGSPROC}} pipeline with the flag `withbackgroundmodel=yes'). 
After inspecting the $2^{\rm nd}$-order spectra, we decided not to use 
them as they were highly affected by the background.
We stacked the $1^{\rm st}$-order RGS 1 and 2 spectra from all observations 
with {\scriptsize{RGSCOMBINE}} and the EPIC-pn, MOS\,1 and MOS\,2 spectra, 
using {\scriptsize{EPICSPECCOMBINE}}.
The stacking provided 4 time-averaged high signal-to-noise spectra for RGS,
MOS\,1, MOS\,2, and pn detectors.
%%%, which are shown in Fig.\,\ref{Fig:Plot_XMM_wind_bestfits}.

%%%The RGS extraction region includes along the dispersion direction several other X-ray sources, all significantly fainter that ULX-1 (see Fig.\,\ref{Fig:Plot_EPIC_image}). The brightest and closest of all these sources is XMMU J004710.0-204708. In the Appendix \ref{sec:appendix_source_x2} we show that the RGS spectra, and in particular the spectral lines, are not significantly affected by the presence of this source.

All XMM-\textit{Newton} spectra were grouped 
in channels of at least 1/3 of the spectral resolution, for optimal binning 
and to avoid over-sampling, and at least 25 counts per bin, using SAS task {\scriptsize{SPECGROUP}}. 
This has also the advantage to smooth the background spectra in the energy range with low statistics,
avoiding narrow spurious features introduced by the background subtraction.
This also enabled us to check our results with the $\chi^2$ statistics.
The stacked spectra have many counts with the 
binning affecting only the spectral
range at the rather low and high energies
(outside the 0.6-1.7 keV RGS band and above 4 keV in EPIC spectra), 
where line detection is not crucial.
We found no significant effect onto our line or continuum modelling 
by decreasing the binning to just 1/3 of the spectral resolution.

\subsubsection{Optical Monitor}
\label{sec:om_reduction}

We used OM data to search for a possible optical/UV counterpart to NGC 247 ULX-1. 
To increase the signal-to-noise ration we stacked all the internally aligned full-frame 
sky images per filter and per observation, using the {\scriptsize{SAS}} tool {\scriptsize{OMMOSAIC}}.
%%%and setting the minimum correlation coefficient to 0.5 \textcolor{black}{(@Anto: What does this do?)}.
Each observation contains at least one image in one of these filters: V, UVW1, UM2 and UVW2 
and the final total exposures corresponded to 105 ks, 105 ks, 115 ks and 220 ks, respectively.
We ran the {\scriptsize{OMDETECT}} task on these stacked images with a limit on the detection 
threshold of $2\,\sigma$.
%%% and adopting the background method estimation of \textcolor{black}{Bickel et al. 2002}. 
At the position of the ULX (Section \ref{sec:data_reduction_epic}), no source was detected in any of the filters. 
This is unsurprising since the ULX region is very close to a bright association of OB stars, 
which makes such a detection challenging. To derive an upper limit for the ULX emission in these bands, 
we computed the total background rate for a circular region of 6'' around the  
position of the ULX. 
This provided $3\,\sigma$ upper limits for the ULX flux, 
which are comparable to previous measurements obtained by \citet{Feng2016}
using deep observations with \textit{Hubble Space Telescope} (HST, UV-optical fluxes 
$\sim10^{-14}$ erg s$^{-1}$ cm$^{-2}$).
In the far-UV the OM flux upper limits are slightly below the HST detections, suggesting long-term 
flux variability and that a substantial fraction of the UV flux originates within the accretion disc 
rather than the stellar companion (see Fig. \ref{Fig:Plot_SED_Balance}, top panel).

\begin{figure}
  \includegraphics[width=1\columnwidth, angle=0]{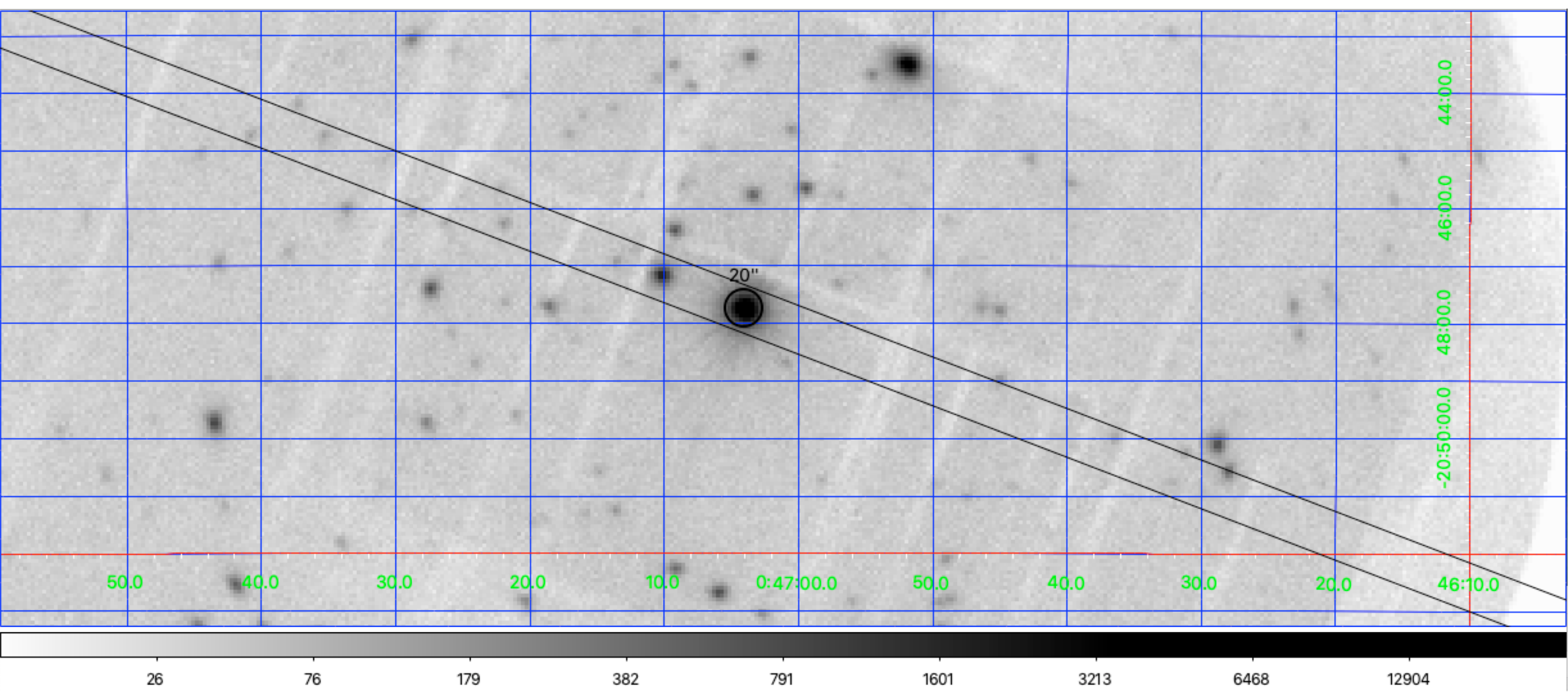}
  \includegraphics[width=1\columnwidth, angle=0]{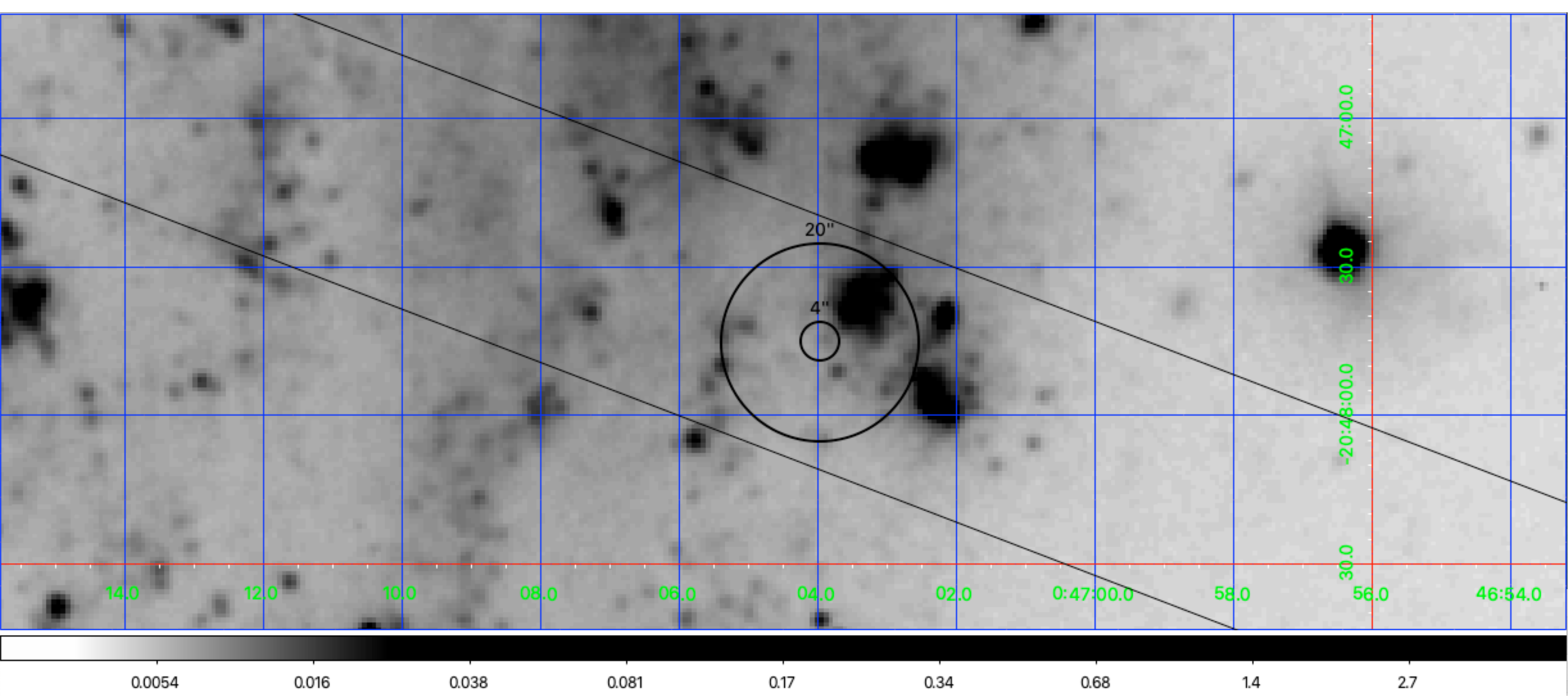}
   \caption{XMM-\textit{Newton} image of the NGC 247 field obtained by combining all the data 
                available for EPIC-pn and MOS\,1,2 zooming onto the ULX-1 region (top panel). 
                The black strip and circle indicate the RGS and EPIC extraction regions, respectively.                
                The bottom panel shows the time-averaged
                image obtained by stacking all the data from the Optical Monitor.
                A small circle with 4" radius shows the X-ray source centroid.}
   \label{Fig:Plot_EPIC_image}
   \vspace{-0.3cm}
\end{figure}

A final OM image obtained by stacking the data from all the filters of all the observations is shown
in Fig.\,\ref{Fig:Plot_EPIC_image}, bottom panel, where the bright association of OB stars can be 
seen on the right side of the X-ray source centroid.

%%%Optical monitor: A summary of the results is shown in Table 1.
%%%\textcolor{black}{\noindent See Anto's in subdirectory: Other\_info/om\_analysis.pdf}\\

\begin{figure*}
  \includegraphics[width=1.33\columnwidth, angle=0]{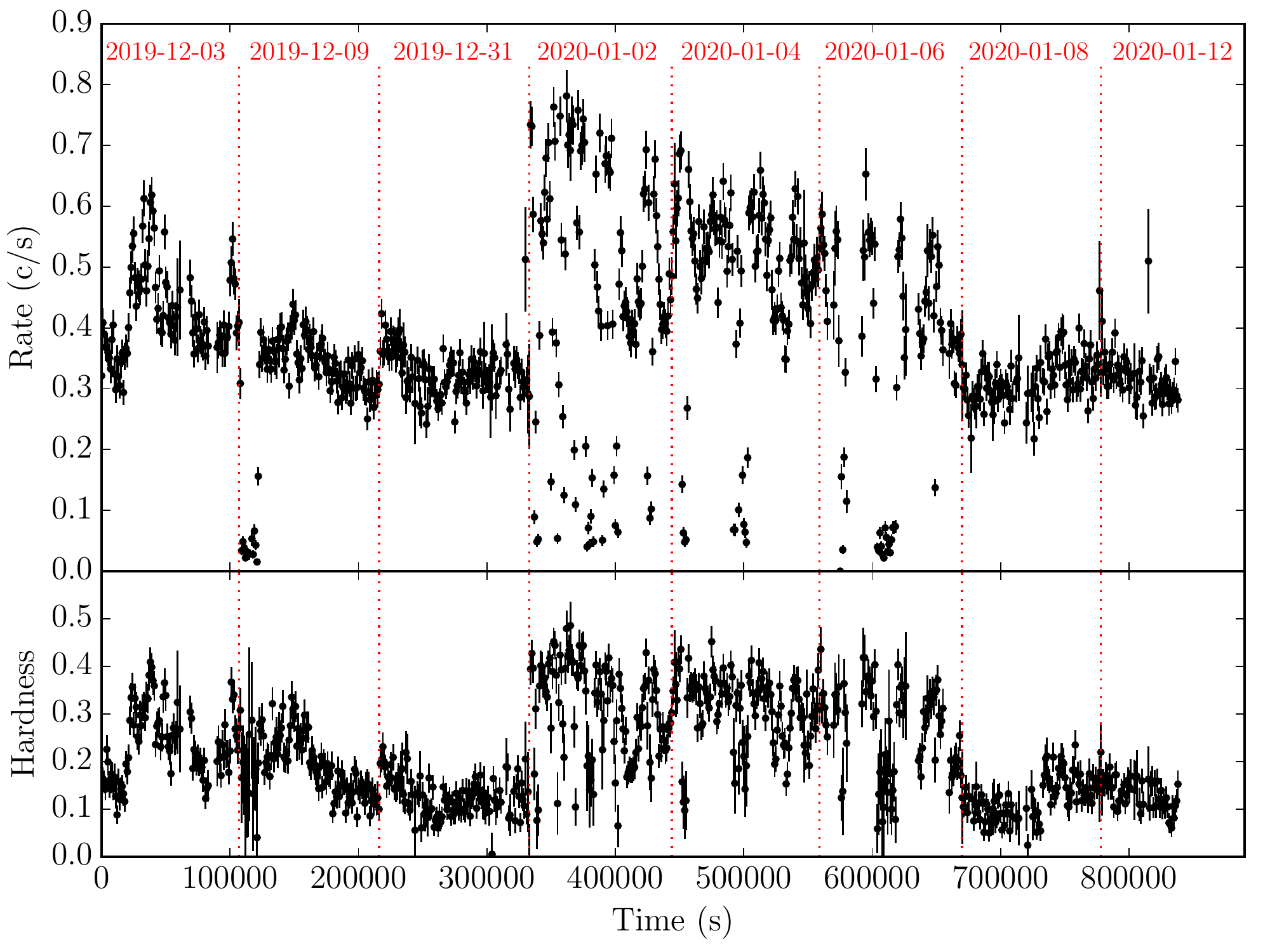}
  \includegraphics[width=0.66\columnwidth, angle=0]{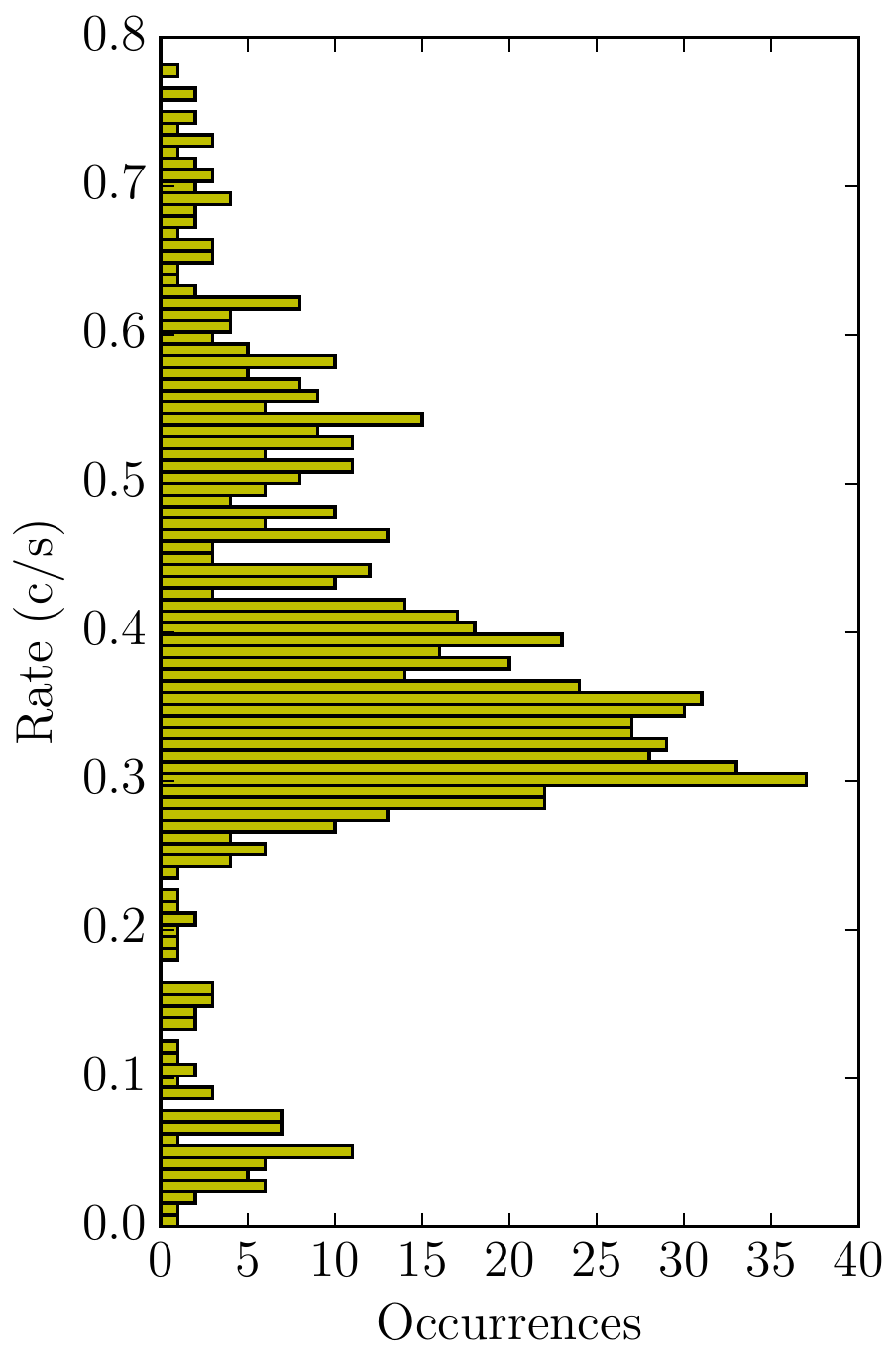}
   \vspace{-0.2cm}
   \caption{Top left panel: cumulated XMM-\textit{Newton} (0.3--10 keV) pn lightcurve 
                 of NGC 247 ULX-1 
                 with the individual observations separated by vertical dotted lines. 
                 Time bin size is 1 ks. \textcolor{black}{Obsid 3-to-7 are separated by 20-60\,ks each.}
                 Bottom left panel: hardness ratio (H/(H+S)) estimated from the lightcurves
                 in the soft (0.3--1 keV) and hard (1--10 keV) X-ray bands.
                %%% Strong flux dips associated to lower hardness are seen on time scales of less than 1 ks.
                 Right panel: count rate histogram.}
   \label{Fig:Plot_XMM_lc}
   \vspace{-0.3cm}
\end{figure*}

\subsubsection{Data investigation} %%%\textcolor{black}{(better in the discussion?)}
\label{sec:data_investigation}

The XMM-\textit{Newton} 0.3--10 keV lightcurve shows a strong dipping behaviour 
with the source flux approaching zero c/s during time interval of less than 15 ks, as found by
\citet{Feng2016}, see Fig.\,\ref{Fig:Plot_XMM_lc} 
(top-left panel). The dips have variable duration, with the shortest ones being of
a few hundred seconds, which was estimated extracting finer lightcurves (D'A\`i et al. in prep).
During the dips, the flux drops by an order of magnitude and then returns to the previous level,
which makes it difficult to believe it is due to an intrinsic flux change rather than to a temporary 
obscuration phenomenon. This behaviour causes the multiple peaks present in the histogram of the 
count rates (Fig.\,\ref{Fig:Plot_XMM_lc}, right panel), which disagree with a single-peaked log-normal trend.

The hardness ratio decreases during the dips (Fig.\,\ref{Fig:Plot_XMM_lc}, bottom-left panel),
which is very similar to the soft source NGC 55 ULX-1 \citep{Pinto2017}. This was
interpreted as evidence of temporary obscuration of the inner hot regions from a clumpy disc wind
({\it e.g.}, \citealt{Stobbart2006}).

The lightcurve also shows that NGC 247 ULX-1 undergoes a long-term flux variability with  
observations 4-to-6 exhibiting significantly higher fluxes. Importantly, the \textcolor{black}{frequency
of the dips} was higher during these observations. 

In order to evaluate the source variability during each observation, we calculated
the fractional excess variance of the EPIC-pn lightcurve of each observation and the root-mean-square (RMS),
following standard formulae (see, {\it e.g.}, \citealt{Nandra1997}, \citealt{Vaughan2003}, and 
\citealt{Allevato2013}). We adopted time bins of 1 ks and, as time length, the duration of each exposure 
($\sim100$ ks, with the exception of obsid:0844860801).
The computed RMS values are reported in Table\,\ref{table:continuum_allabs} and 
range from about 10 to 55\,\%.
%%%  the highest values provided by the dipping observations.
%%%FP: This is the fractional variability rather than the RMS, right?

The EPIC spectra of the individual observations in Fig.\,\ref{Fig:Plot_EPIC_spectra} 
show a variability pattern that is common to ULXs, with the harder band ($>1$\,keV) 
exhibiting the largest variation (see, {\it e.g.}, \citealt{Middleton2015a}, \citealt{Brightman2016},
and \citealt{Walton2018c}).
%%%The spectra seem to show pivoting which is evidence of multiple ongoing variability processes
%%%such as the long-term ($\gtrsim10$ ks) smooth flux changes and the short term ($\lesssim10$ ks) 
%%%flux dips. 
A thorough analysis of the variability pattern involving a careful sampling of 
time interval with similar flux and hardness ratio, and the study of the power density spectra,
will be done in two separate papers 
(D'A\`i et al. in prep, \citealt{Alston2021}).
%%%Here we focussed on the search for outflows in the high-statistics time-averaged spectrum
%%%and for variations in the spectral residuals.
 
\section{Spectral analysis}
\label{sec:spectral_analysis}
 
%\textcolor{black}{Plan 1) 1keV variation in CCD spectra, 2) Time-averaged XMM continuum; 3) identify Gaussians in RGS, 4) Wind Models (CIE, SED+PIE) + Monte Carlo}

\textcolor{black}{In this section we present the spectral analysis of NGC 247 ULX-1. We first show
the time evolution of the main spectral residuals around 1 keV through the modelling of the spectral
continuum in EPIC spectra from different observations (see Sect. \ref{sec:residuals_variability}).
Then we will perform a thorough analysis of the high-statistics, time-averaged, stacked spectrum
in order to identify the lines in the RGS (Sect. \ref{sec:baseline_continuum} and \ref{sec:line_search}), 
to build the spectral energy distribution (SED),
and to use physical plasma models for the wind detection and modelling
(Sect. \ref{sec:cie_modelling} and \ref{sec:photoionisation_modelling}).
The final best-fit models are shown in Sect. \ref{sec:model_comparison} and the statistical significance of our findings in Sect. \ref{sec:mc_simulations}.}
 
ULX spectra require up to three components to obtain
a satisfactory description of the spectral continuum. 
Two blackbody-like components are often used to model the soft (0.3--1 keV) and 
hard (1--10 keV) X-ray energy bands (see, {\it e.g.}, \citealt{Stobbart2006}, \citealt{Pintore2015}). 
The availability of high-statistics spectra and broadband
data reveals the presence of a third harder component, which
dominates the continuum above 10 keV (see, {\it e.g.}, \citealt{Walton2018a}).
In the framework of super-Eddington accretion, the cooler, soft, component corresponds
to the X-ray emission of the wind and the disc around the spherisation radius.
The hard component refers to the inner accretion flow.
The hard ($>$ 10 keV) tail is either due to Compton scattering in the innermost regions 
or from an accretion column (see, {\it e.g.}, \citealt{Middleton2015a}).
The two hard components, especially the hard tail, are weak in supersoft ULXs.

\subsection{Time evolution of the $\sim$1\,keV residuals}
\label{sec:residuals_variability}

For the modelling of EPIC spectra of individual observations we adopted 
a simple continuum model consisting of two blackbody (\textit{bb}) components, which is often
used as a proxy for more complex models 
(see, {\it e.g.}, \citealt{Walton2014}, \citealt{Pinto2017}, \citealt{Koliopanos2017}, 
and \citealt{Gurpide2021}).
We did not model the hard tail in the individual spectra because it is so weak 
that any model would be highly unconstrained, but for the time-averaged
SED modelling we took it into account (see Sect.\,\ref{sec:baseline_continuum}).
The emission components are corrected for absorption by the Galactic interstellar
medium and the circumstellar medium near the ULX using the \textit{hot} model in 
{\scriptsize{SPEX}} with a low temperature ($kT=0.2$\,eV, {\it e.g.} \citealt{Pinto2013}),
at which the gas is neutral.
In the spectral fits, we coupled the column density of the \textit{hot} model 
across all observations 
as it is unlikely that the amount of neutral gas along our line of sight 
towards the ULX would change on time scales of a few days.
In some Galactic X-ray binaries (XRB) variable obscuration was found, but it is not clear 
whether these findings are related to ionised gas ({\it i.e.} winds) or uncertainties in the continuum
rather than neutral gas (see, {\it e.g.}, \citealt{Miller2009}, \citealt{Walton2017}).
%%%Moreover, a preliminary spectral fit performed with free column densities 
%%%in the three spectra using the \textit{bb}$_{1,2}$ model
%%%yields broadly consistent column densities.

\begin{figure*}
  \includegraphics[width=1.8\columnwidth, height=11cm, angle=0]{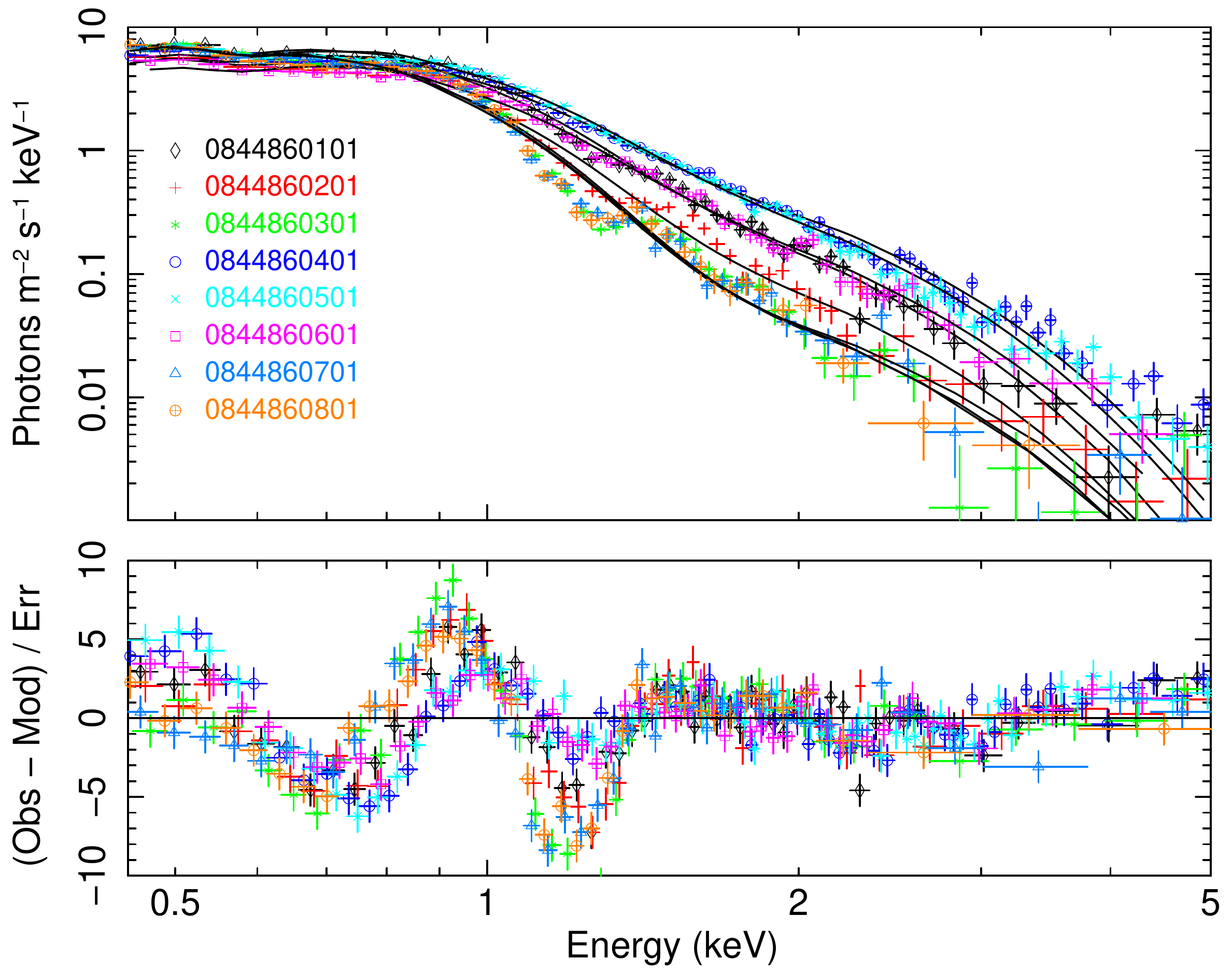}
\vspace{-0.1cm}
   \caption{XMM-\textit{Newton} EPIC-pn spectra of NGC 247 ULX-1.
   The top panel shows the pn spectra of the individual observations
   with overlaid the 2-blackbody continuum models. 
   The bottom panel shows the corresponding residuals. 
   Both the emission feature below 1 keV and the absorption feature above 1 keV
   vary in centroid and strength according to the continuum flux and shape.
   This is qualitatively similar to the intermediate-hard 
   source NGC 1313 ULX-1 and the soft source NGC 55 ULX-1 \citep{Middleton2015b, Pinto2017}.}
   \label{Fig:Plot_EPIC_spectra}
\vspace{-0.5cm}
\end{figure*}

We simultaneously applied the $hot\,(bb+bb)$ continuum model to the EPIC MOS 1,2 and pn spectra
of \textcolor{black}{all} eight observations. 
The results are shown in Fig.\,\ref{Fig:Plot_EPIC_spectra} and Table\,\ref{table:continuum_allabs}.
We obtained an average column density
of $N_{\rm H} = (3.4 \pm 0.1) \times 10^{21} {\rm cm}^{-2}$.
This phenomenological model provides
a good description of the broadband spectra, but the $C$-statistics are high when compared 
to the corresponding degrees of freedom 
($C_{\nu}\sim3-9$) %%% (perhaps quote individual C/d for each obs MOS1,2,pn)
due to the well-known strong and sharp residuals in the form
of absorption and emission features around 1 keV (see Fig.\,\ref{Fig:Plot_EPIC_spectra}). 

\begin{figure*}
  \includegraphics[width=1\columnwidth, angle=0]{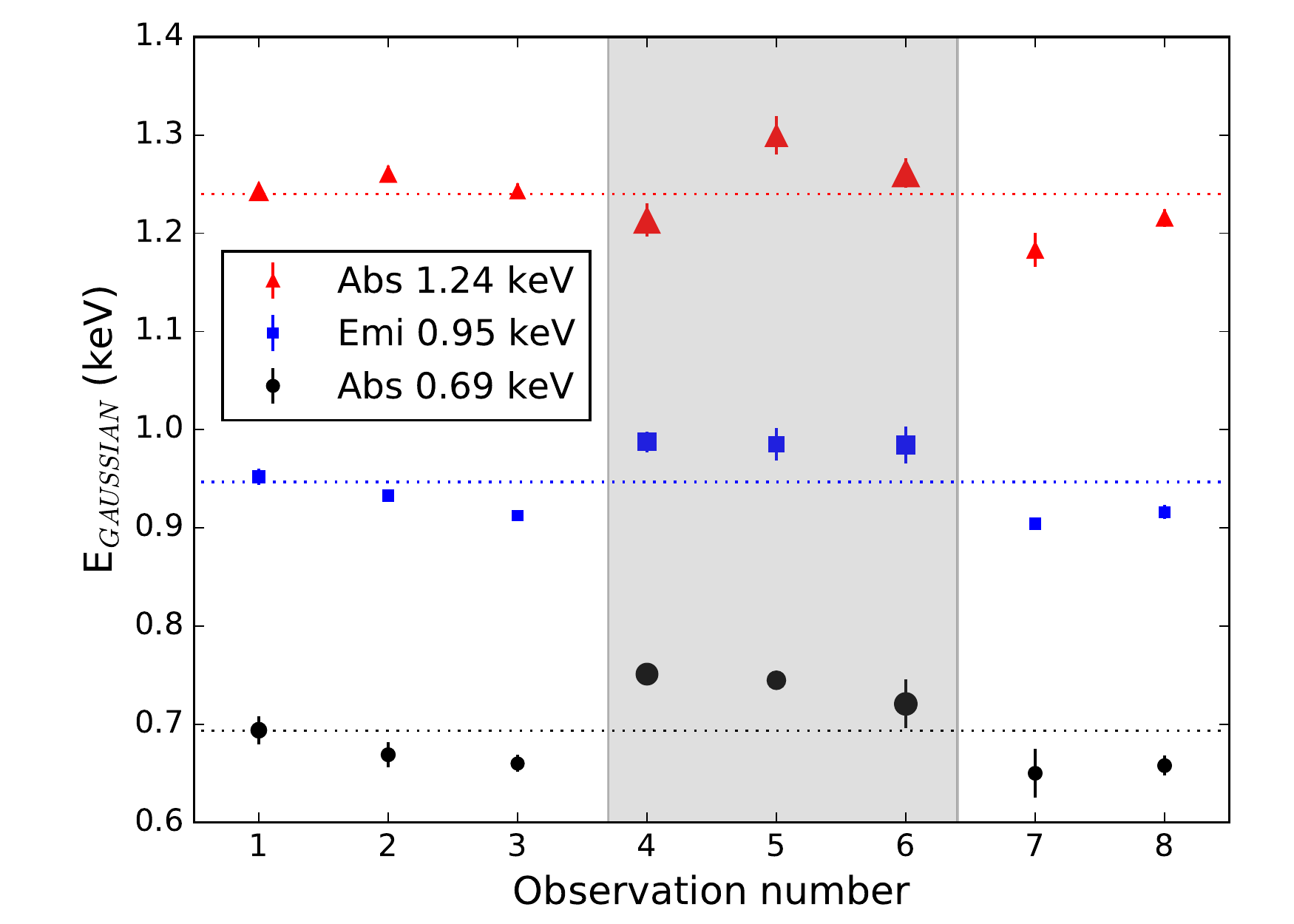}
  \includegraphics[width=1\columnwidth, angle=0]{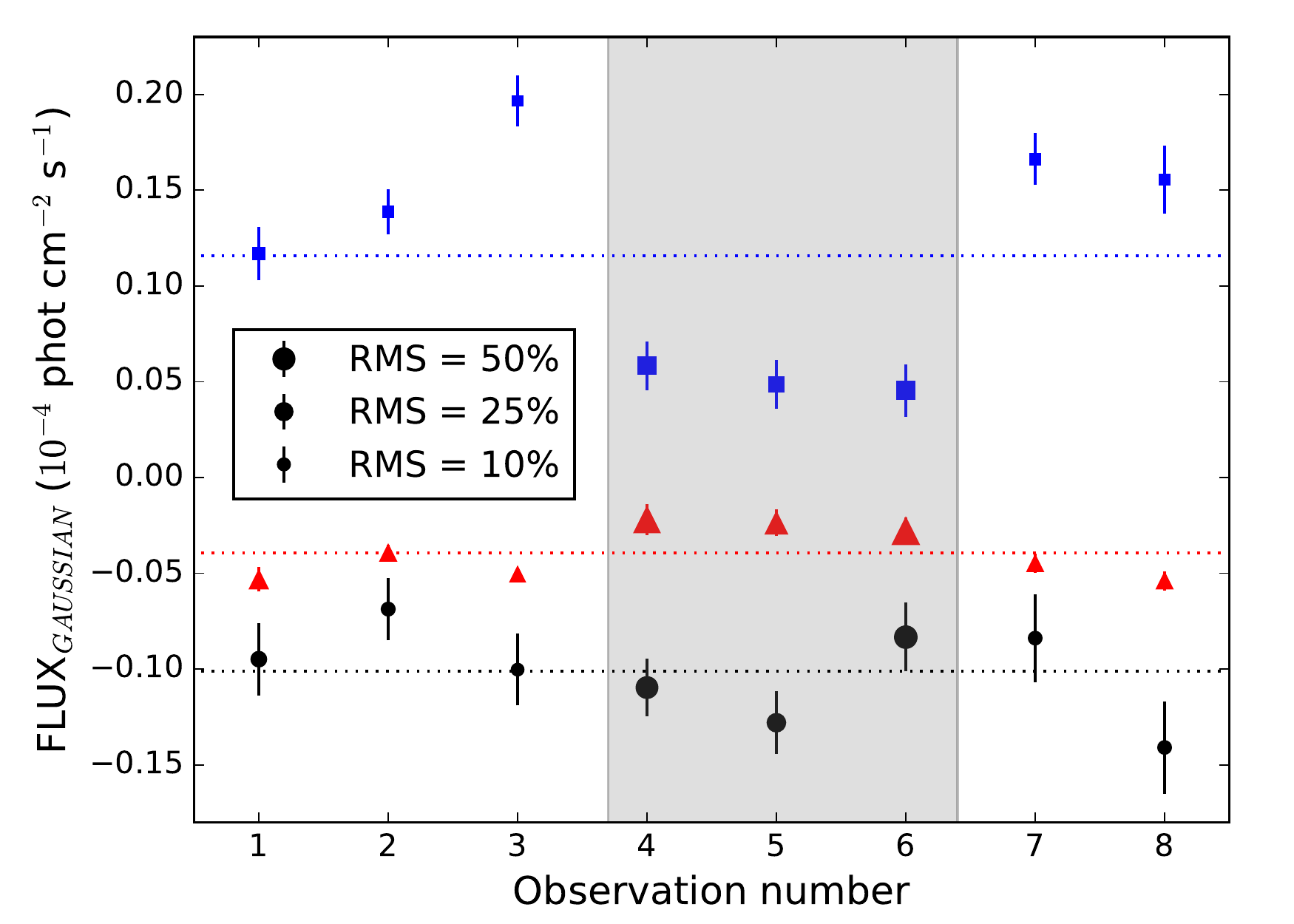}
\vspace{-0.1cm}
   \caption{Evolution of the XMM-\textit{Newton} EPIC spectral residuals around 1 keV.
   The left and the right panels show the energy centroid and normalisation of the Gaussian
   lines used to fit the three main (unresolved) spectral residuals, respectively.
   The shaded grey areas highlight the observations of high spectral variability and dipping.
   `Abs' and `Emi' refer to absorption and emission features, respectively.}
   \label{Fig:Plot_EPIC_spectra_res}
\vspace{-0.2cm}
\end{figure*}

It is possible to understand the nature of the residuals by
tracking their temporal evolution from one observation to another.
In \citet{Middleton2015b} and \citet{Pinto2017}, 
the three main \textcolor{black}{spectral features observed
around 1 keV in the spectra of NGC 1313 ULX-1 and NGC 55 ULX-1 were modelled} 
with a positive (emission) and two negative Gaussians (absorption) lines.
The availability of deeper observations allow us to fit the three
Gaussian components independently, but we fixed the line broadening to zero km/s
({\it i.e.} only instrumental broadening) in order to minimise the degeneracy
that can be produced by the low spectral resolution of EPIC. 
We chose three Gaussian lines as previous work on high-resolution RGS
data identified Fe\,/\,Ne emission lines at 1 keV, {\oviii} absorption lines
around 0.7--0.8 keV, and Fe\,/\,Ne absorption lines above 1 keV (\citealt{Pinto2016nature}). 
\textcolor{black}{Using only one or two Gaussian lines always resulted in significantly 
worse fits during alternative tests.}

The $hot\,(bb+bb) + (gaus+gaus+gaus)$ spectral model improves the fits for all
observations with respect to the $hot\,(bb+bb)$ model. 
In Table\,\ref{table:continuum_allabs}, we report the best-fitting parameters for 
each observation. 
The reduced C-stat are still rather high due to additional residuals that require 
more complex and physical models.
Moreover there is evidence of a weak, broad, hard tail in all spectra above 3 keV,
which cannot be explained with atomic lines (Fig. \ref{Fig:Plot_EPIC_spectra}).

In Fig.\,\ref{Fig:Plot_EPIC_spectra_res}, we compare the energy centroids (left panel)
and the fluxes (right panel) as measured for the three Gaussian lines in the EPIC spectra
of the eight observations. The point size was coded according to the value of their RMS 
estimate in Sect.\,\ref{sec:data_investigation}. Both the energies and fluxes of the Gaussians
lines vary with the time, showing a higher blueshift and lower flux (in absolute value)
during the dipping observations with enhanced variability.

\subsection{Time-averaged continuum: spectral modelling}
\label{sec:baseline_continuum}

The RGS spectra of the individual observations do not provide statistics sufficient
to detect and resolve the spectral residuals with high significance.
The 750\,ks RGS 1+2 stacked spectrum instead has a much higher quality and 
enables line detection, despite the low source flux.

\begin{figure}
  \includegraphics[width=\columnwidth, angle=0]{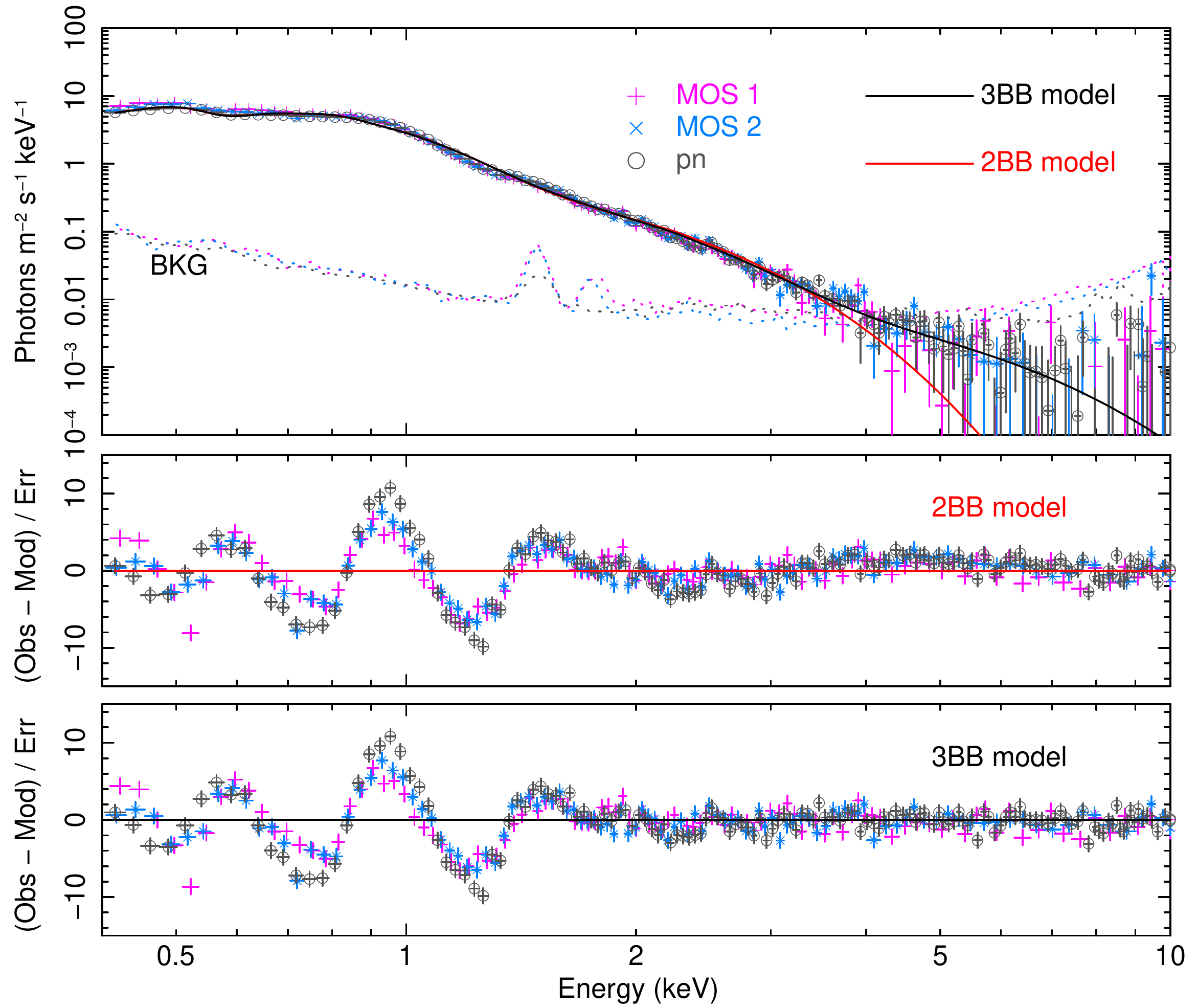}
  \vspace{-0.3cm}
   \caption{Time-averaged XMM-\textit{Newton} EPIC-pn and MOS 1,2 spectra.
                 Overlaid are two alternative continuum models consisting of two (red line) 
                 and three (black line) blackbody components.}
   \label{Fig:NGC247_RGS_EPIC_all_YLOG}
  \vspace{-0.2cm}
\end{figure}

\begin{figure*}
  \includegraphics[width=1.8\columnwidth, height=9.5cm, angle=0]{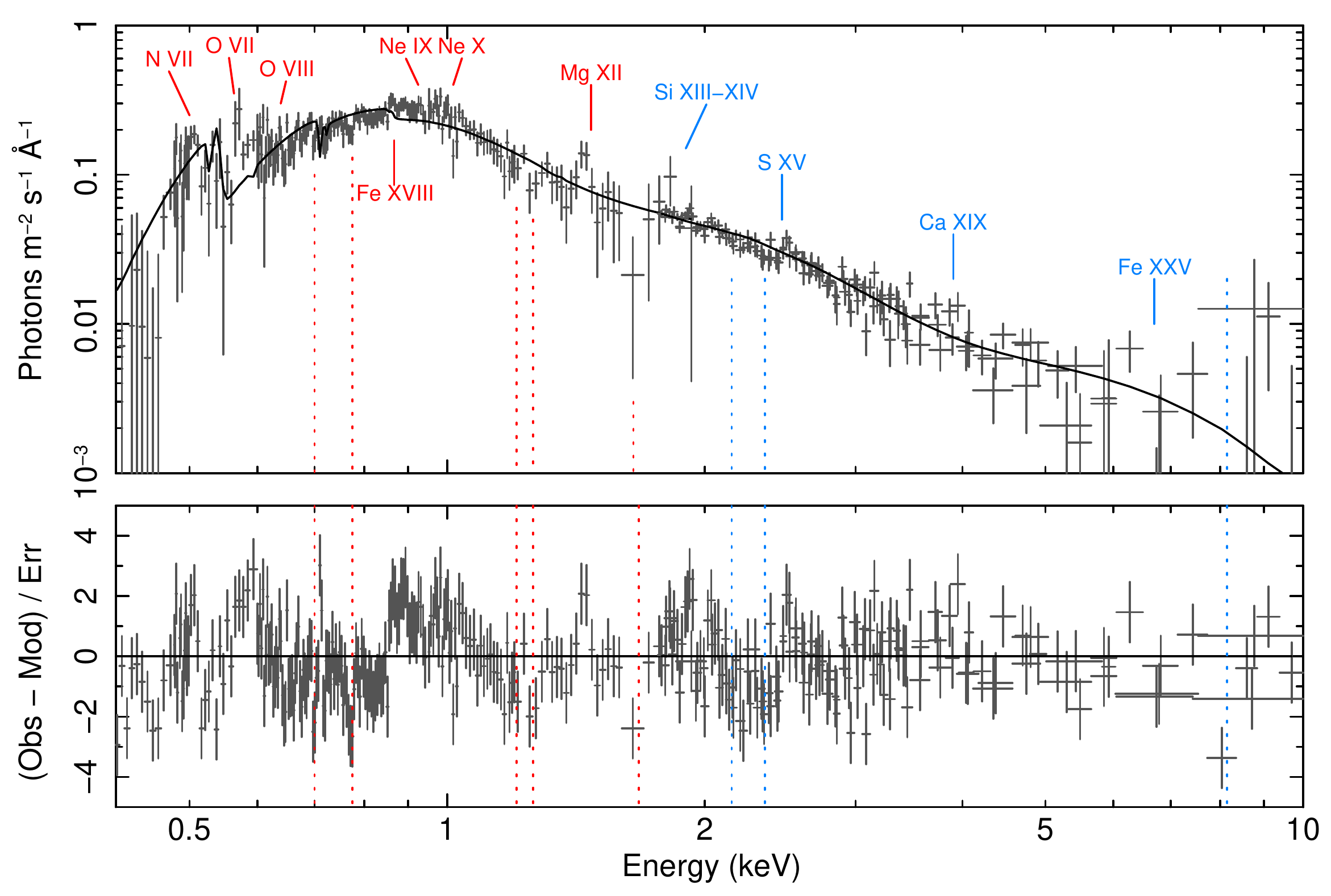}
  \vspace{-0.2cm}
   \caption{Top panel: time-averaged XMM-\textit{Newton} RGS (0.33$-$2 keV) 
                 and EPIC MOS\,1,2-pn (1.77$-$10 keV) spectra.
                 Overlaid is the baseline 3-blackbody continuum model (solid black line,
                 see Table \ref{table:continuum_timeavg}).
                 The bottom panel shows the residuals calculated with respect to the continuum 
                 model.
                 The rest-frame energies of the most common strong transitions in the X-ray band 
                 (red for RGS band and blue for EPIC) 
                 and the absorption features (dotted lines) are also shown.
                 All spectra were rebinned for displaying purposes.
                 }
   \label{Fig:Plot_XMM_wind_bestfits}
  \vspace{-0.4cm}
\end{figure*}

\begin{figure}
  \includegraphics[width=\columnwidth, angle=0]{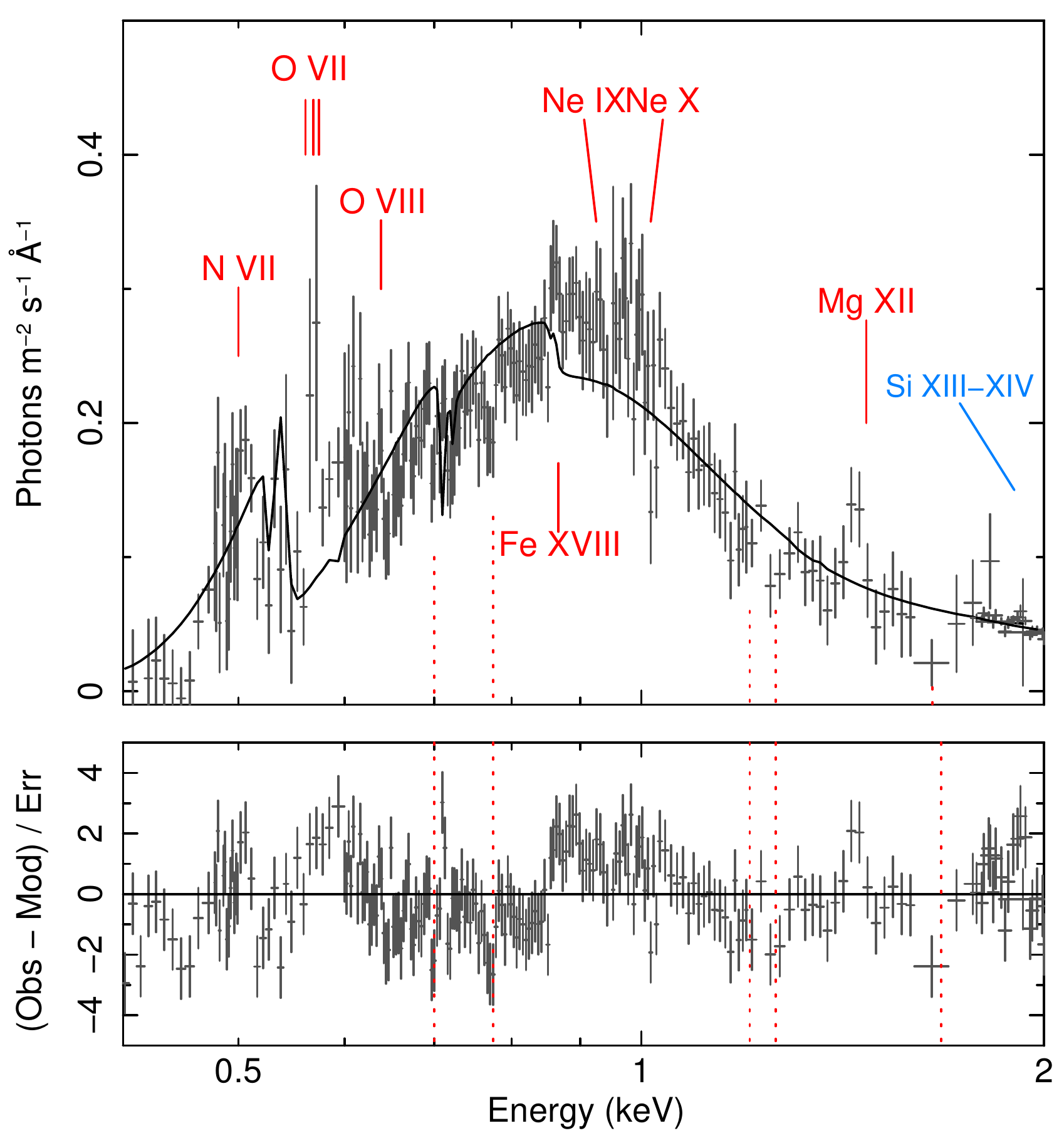}
  \vspace{-0.6cm}
   \caption{Time-averaged XMM-\textit{Newton} RGS (0.33$-$2 keV) and EPIC-pn (1.77$-$10 keV) 
                 spectra.
                 Overlaid is the baseline continuum model.
                This is a zoom of Fig.\,\ref{Fig:Plot_XMM_wind_bestfits} onto the RGS data.}
   \label{Fig:Plot_XMM_wind_bestfits_zoom}
  \vspace{-0.4cm}
\end{figure}

We simultaneously fitted the time-averaged stacked EPIC MOS 1,2 and pn, 
and the RGS spectra using the absorbed double-blackbody continuum model ($hot\,(bb+bb)$)
adopted for the spectra of the individual observations. The hard tail above 3 keV is more
evident in the stacked data and, therefore, we accounted for it using a third, hotter 
($kT\sim1$ keV), blackbody as to mimic additional hard X-ray photons 
down-scattered through the disc photosphere (and/or the wind).
The three-blackbody model brings the C-stat from 4671 down to 4488 for four additional 
degrees of freedom. In all fits the parameters of the blackbody and the ISM absorber were 
coupled among the EPIC and RGS models.
We left the overall normalisations of the MOS 1,2 and RGS models free to vary with respect to pn
in order to account for the typical 5-10\% cross-calibration uncertainties among their effective areas.
Details on the spectral fits for both models are reported in Table\,\ref{table:continuum_timeavg}
and Fig.\,\ref{Fig:NGC247_RGS_EPIC_all_YLOG}. 
In order to avoid a crowded plot, we did not include the more noisy RGS spectrum in this plot, while 
it is shown later in Figs.\,\ref{Fig:Plot_XMM_wind_bestfits} and \ref{Fig:Plot_XMM_wind_bestfits_zoom}
where the EPIC data below 1.77 keV was ignored.

The blackbody models used so far are the simplest available and allowed us to constrain the parameters. 
We tested various combination of two-components models to possibly improve the description of the spectral continuum before accounting for narrow features. These consisted of the cool blackbody component plus 
either a disc blackbody ($dbb$) or a disc blackbody modified by coherent 
Compton scattering ($mbb$) or Comptonisation ($comt$). These 
did not provide improvements with respect to the three blackbody model. Similar results were provided by
more complex three-component continuum models which anyway
over-fit the data and lead to degeneracy among model parameters due to the weak hard 
($>1$ keV) continuum. Alternatively, one could use the common blackbody plus powerlaw
emission model. However, the powerlaw would be very steep with a consequent unphysical divergence
in the soft band, which would badly affect the ionisation balance calculation 
(see, {\it e.g.}, \citealt{Pinto2020a}).

\begin{table}
\caption{Time-averaged EPIC+RGS spectral fits.}  
\label{table:continuum_timeavg}     
\renewcommand{\arraystretch}{1.1}
 \small\addtolength{\tabcolsep}{0pt}
 
\scalebox{0.9}{%
\hskip-0.0cm\begin{tabular}{@{} l l l l}     
\hline
Parameter                 & Units                                 & $hot\,(bb+bb)$         & $hot\,(bb+bb+bb)$  \\  
\hline  
$Area,bb_1$             & $10^{19}$cm$^2$             & $3.4\pm0.3$            & $4.5\pm0.3$            \\  
$Area,bb_2$             & $10^{16}$cm$^2$             & $1.0\pm0.1$            & $1.7\pm0.2$            \\  
$Area,bb_3$             & $10^{13}$cm$^2$             &   --                            & $1.7\pm0.5$            \\  
$kT,bb_1$                 &  keV                                  & $0.120\pm0.001$    & $0.116\pm0.001$    \\  
$kT,bb_2$                 &  keV                                  & $0.382\pm0.003$    & $0.342\pm0.005$    \\  
$kT,bb_3$                 &  keV                                  &  --                             & $1.05\pm0.07$        \\  
L$_{X,tot}$                & $10^{39}$erg/s                 & $5.3\pm0.5$            & $6.2\pm0.5$             \\  
$N_{\rm H}$              & $10^{21} {\rm cm}^{-2}$   & $3.64\pm0.05$        & $3.84\pm0.05$         \\  
C-stat/d.o.f.               &                                          & $4671/1136$            & $4488/1132$           \\  
\hline                                                                                                                
\end{tabular}}

$L_{X\,(0.3-10\,\rm keV)}$ luminosities are calculated assuming a 
distance of 3.3\,Mpc and are corrected for absorption (or de-absorbed).
The best-fit three-blackbody model and the spectra are shown in 
Fig.\,\ref{Fig:Plot_XMM_wind_bestfits}. Both models are shown in the SED
modelling in Fig.\,\ref{Fig:Plot_SED_Balance}.
\end{table}

\subsection{Time-averaged continuum: Gaussian line scan}
\label{sec:line_search}

In Fig.\,\ref{Fig:Plot_XMM_wind_bestfits}, we show the stacked RGS and EPIC spectra,
indicating the dominant \textcolor{black}{H-/He-like} transitions of the X-ray band
\textcolor{black}{often found in the spectra of X-ray binaries}. A zoom over the RGS
spectrum with ad-hoc linear Y-axis can be found in Fig.\,\ref{Fig:Plot_XMM_wind_bestfits_zoom}.
The RGS exhibits substantial residuals at the same energies as the EPIC residuals but
resolves them in a structure of lines, although the former have lower count rates. 
Strong emission-like features appear near the transition energies of the most
relevant neon \textcolor{black}{K and iron L} lines. Additional features may be related to {\nvii} and {\oviiviii},
although the background starts to be important in the RGS below 0.6 keV.
Some possible absorption-like features are indicated with vertical dotted lines.
The very good agreement between the positions of the RGS, MOS 1,2 and pn residuals rules out 
instrumental dominant features.

Following the approach used in \citet{Pinto2016nature}, we searched for narrow spectral features
by scanning the spectra with Gaussian lines. We adopted a logarithmic grid of 1000 points
with energies between 0.3 (41\,{\AA}) and 10 keV (1.24\,{\AA}).
This choice provided a spacing that is comparable to the RGS and EPIC resolving power 
in the energy range we are investigating 
($R_{\, \rm RGS}\sim100-500$ and $R_{\, \rm EPIC}\sim20-60$). 
We tested line widths ($\sigma_G$=FWHM/2.355) of 100, 250, 500
and 1000 km/s, which are comparable to the RGS resolution. 
At each energy we recorded the $\Delta C$ improvement to the best-fit continuum model
and expressed the significance as the square root of the $\Delta C$.
This provides the maximum significance for each line 
(as it neglects the number of trials). 
We multiplied \textcolor{black}{$\sqrt{\Delta C}$ by the sign of the Gaussian normalisation}
to distinguish between emission and absorption lines.

%%%\vspace{0.3cm}

\begin{figure*}
  \includegraphics[width=1.7\columnwidth, height=9cm, angle=0]{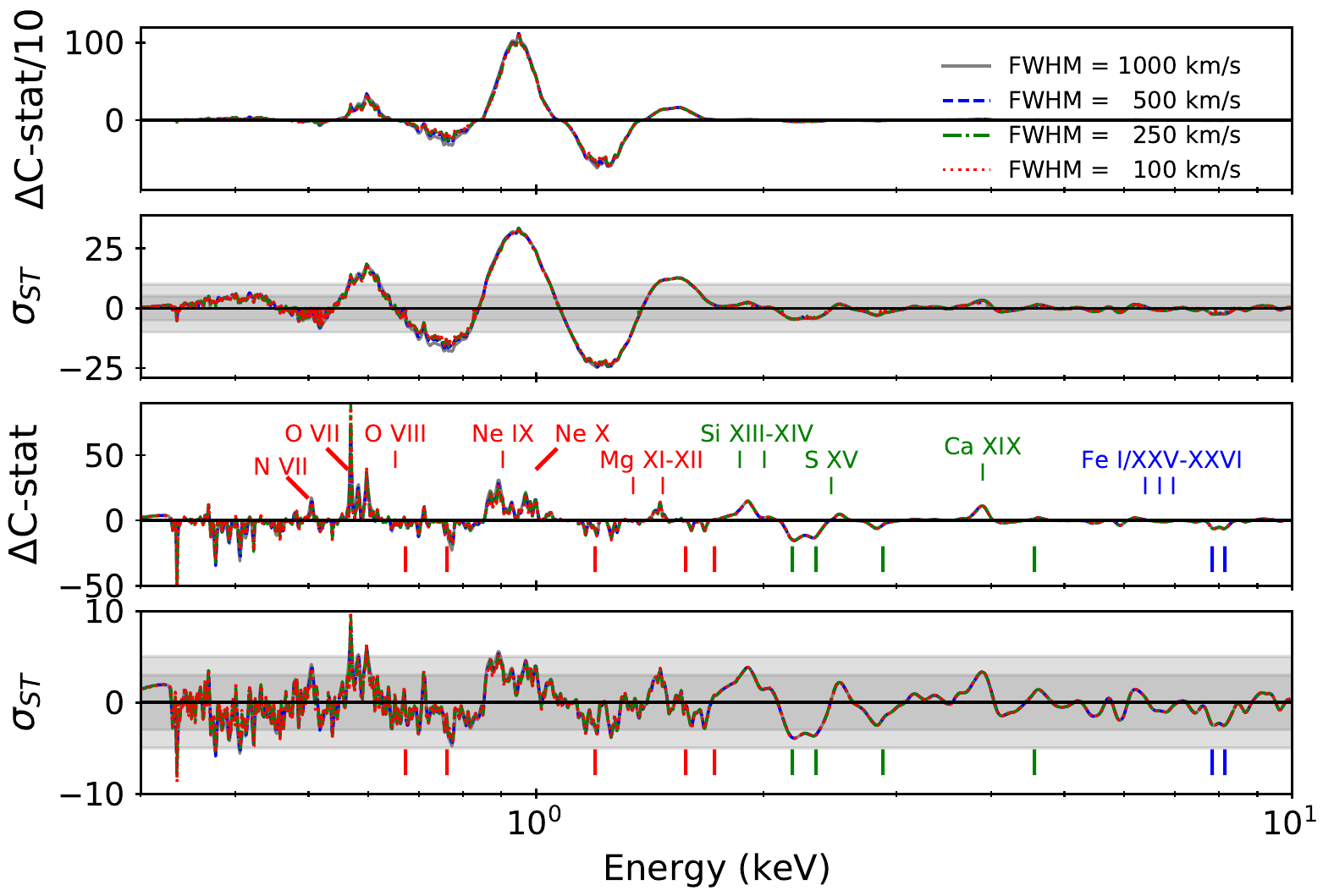}
      \vspace{-0.1cm}
   \caption{Gaussian line scan performed on the time-averaged XMM-\textit{Newton} EPIC 
                 and RGS spectra of NGC 247 ULX-1. The top panels show the case when the EPIC 
                 spectra are used throughout the whole 0.3--10 keV band, while the bottom two panels 
                 show the results obtained with RGS between 0.33$-$2 keV 
                 and EPIC-pn from 1.77$-$10 keV (see also Fig.\,\ref{Fig:Plot_XMM_wind_bestfits}). 
                 The results for four different line widths are shown.
                 \textcolor{black}{No remarkable difference is observed among the adopted widths.}
                 The single-trial line significance (``$\sigma_{ST}$'') is calculated as square root of the 
                 $\Delta C$ times
                 the sign of the Gaussian normalisation (positive/negative for emission/absorption lines).
                 Labels are red for RGS, green for strong EPIC features and blue for the faint Fe K.
                 \textcolor{black}{The grey shaded areas show the 3 and 5 $\sigma_{ST}$ limits for individual lines.}}
   \label{Fig:Plot_line_search}
      \vspace{-0.3cm}
\end{figure*}

In Fig.\,\ref{Fig:Plot_line_search} we show the results of the line scan obtained for the 
time-averaged stacked RGS+EPIC spectra using the three-blackbody continuum model. 
We performed the line scan in two ways: the first run using all RGS data (0.3--2 keV) 
and EPIC data (0.3--10 keV) and the second one ignoring the EPIC data between 
0.33 and 1.77 keV, where the RGS effective area is well calibrated.
When fitting only RGS in the 0.33--1.77 keV energy band we always fixed
the temperatures of the blackbody components to the best-fit values obtained using the EPIC
data in the whole 0.3--10 keV. This is due to the low count rate of the RGS spectra that limited 
our capability to constrain the overall continuum level and shape
(see \citealt{Pinto2020b}).

The line scan of the EPIC + RGS spectra data picked out the strong emission-like features near
1 keV and other two around 0.6 and 1.5 keV. Broad absorption features were also found around
0.7 and 1.2-1.3 keV as previously done in the spectrum of each observation 
(see Fig. \ref{Fig:Plot_EPIC_spectra_res} and Table \ref{table:continuum_allabs}).

Owing to the low spectral resolution and high count rate of EPIC, the features appear 
very broad ($\sim0.1$ keV) in the EPIC+RGS Gaussian scan preventing us from identifying them.
This becomes easier above 2 keV due to the increasingly higher EPIC spectral resolution.
The Gaussian scan performed using only RGS between 0.33 and 1.77 keV resolved the features 
into a forest of lines. Multiple lines are responsible for the 1 keV emission-like and 
the 1.2-1.3 keV absorption-like features. Interestingly, most absorption features are consistent 
with some Lyman $\alpha$ transitions also seen in emission, if we assume a 
\textcolor{black}{systemic blueshifted absorption of about $0.17c$ (see vertical ticks in Fig.
\ref{Fig:Plot_line_search}).} 
The 0.6 keV features might either be interpreted as a blueshifted 
{\ovii}\,$\alpha$ triplet or redshifted {\ovii}\,$\beta$ + {\oviii} Ly\,$\alpha$ emission lines. 
The emission lines found between 0.9-1 keV are most likely from {Ne\,{\sc ix-x}} and 
{Fe\,{\sc xviii-xxiv}} ions. Alternative interpretations correspond to different velocities of the X-ray 
emitting (and absorbing) plasmas. The use of physical models is necessary to distinguish among 
the several interpretations. The most simple physical models involve the adoption of plasma in 
either collisionally-ionisation equilibrium (CIE) or photoionisation equilibrium (PIE). 

\textcolor{black}{The single-trial line significance (``$\sigma_{ST}$'') of the individual strongest features
is around 5\,$\sigma$, which of course is smaller if we take into account the look-elsewhere effect.
However, plasma models are able to model multiple lines, combining their individual $\Delta C$ improvements 
to the best-fit continuum, and boost the overall significance (see below).}

\subsection{Collisional-ionisation jet modelling}
\label{sec:cie_modelling}

\citet{Pinto2020b} performed automated scan models using either CIE or PIE plasmas. 
This technique prevents the fits from getting stuck in local minima, 
although is computationally expensive (lasting a few hours on one CPU).
%%%It also provides a proxy for the absorption (emission) measure distribution 
%%%if applied to line absorption (emission) models.

%%%\textcolor{red}{BREAK: start with NGC 247, move SS 433 to Appendix. Change SS part saying that in the appendix we have also tested SS for corroborating our code.}

\subsubsection{Collisionally-ionised emitting gas}
\label{sec:cie_ngc247}

Following \citet{Kosec2018b} and \citet{Pinto2020b}, 
we performed a multidimensional automated scan with an emission model
that assumes collisional ionisation equilibrium ($cie$ model in {\scriptsize{SPEX}}). 
We adopted a logarithmic
grid of temperatures between 0.1 and 5 keV (50 points), and line-of-sight velocities, $v_{\rm LOS}$, 
between $-0.3c$ (blueshifted jet) and $+0.3c$ (redshifted jet, with steps of 500 km/s). 
We tested several values of velocity dispersion (from 100 to 10000 km/s), 
finding comparable results as already shown by the Gaussian line scan in Fig.\,\ref{Fig:Plot_line_search},
with the best fit achieved at  $v_{\rm RMS}\sim3000$ km/s.
Abundances were chosen to be Solar (to limit the computing time) 
and the emission measure 
$EM=n_{\rm e}\,n_{\rm H}\,V$ was the only free parameter of the CIE in the spectral fit.

\begin{figure*}
   \includegraphics[width=0.95\columnwidth, angle=0]{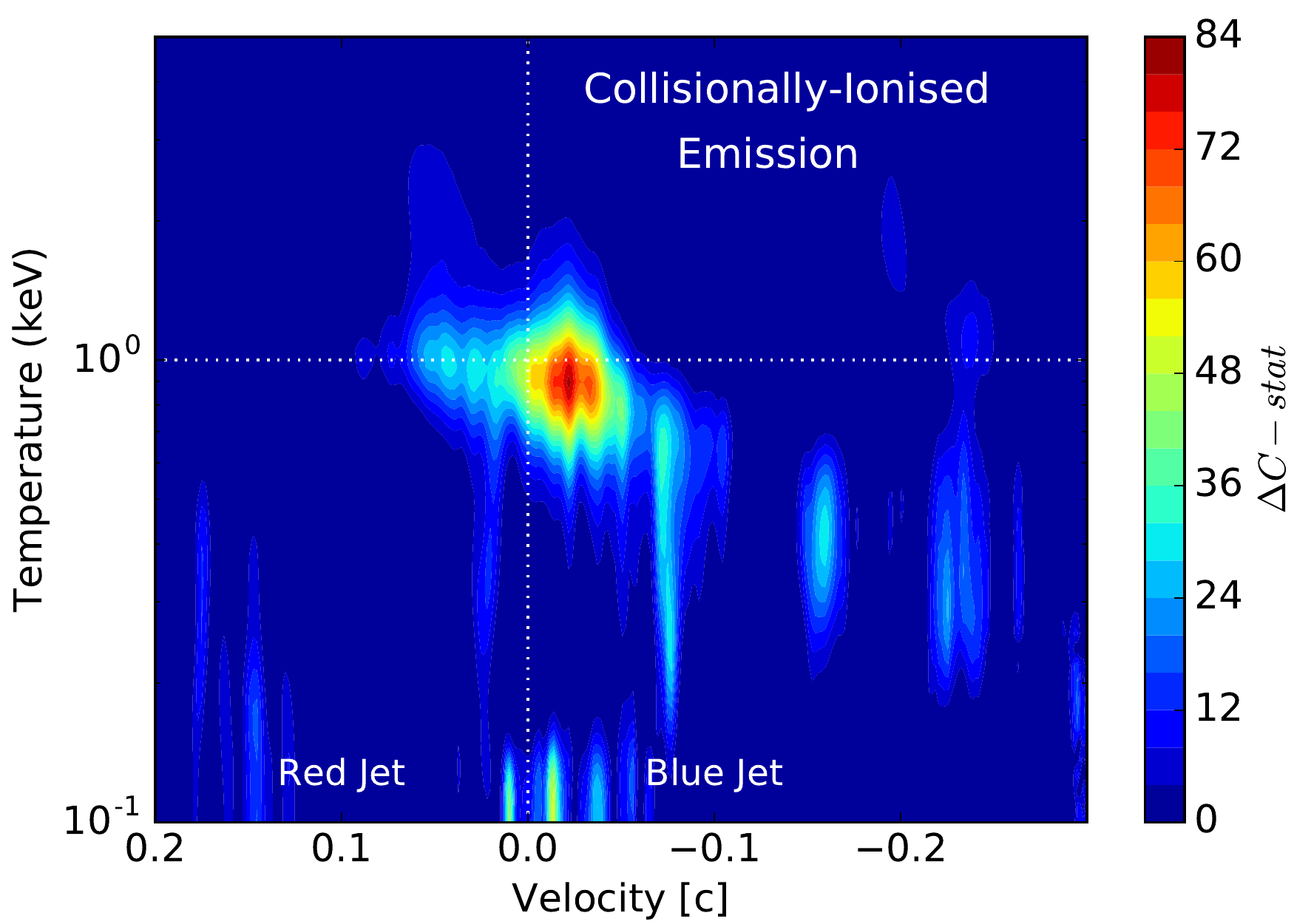}
   \includegraphics[width=0.95\columnwidth, angle=0]{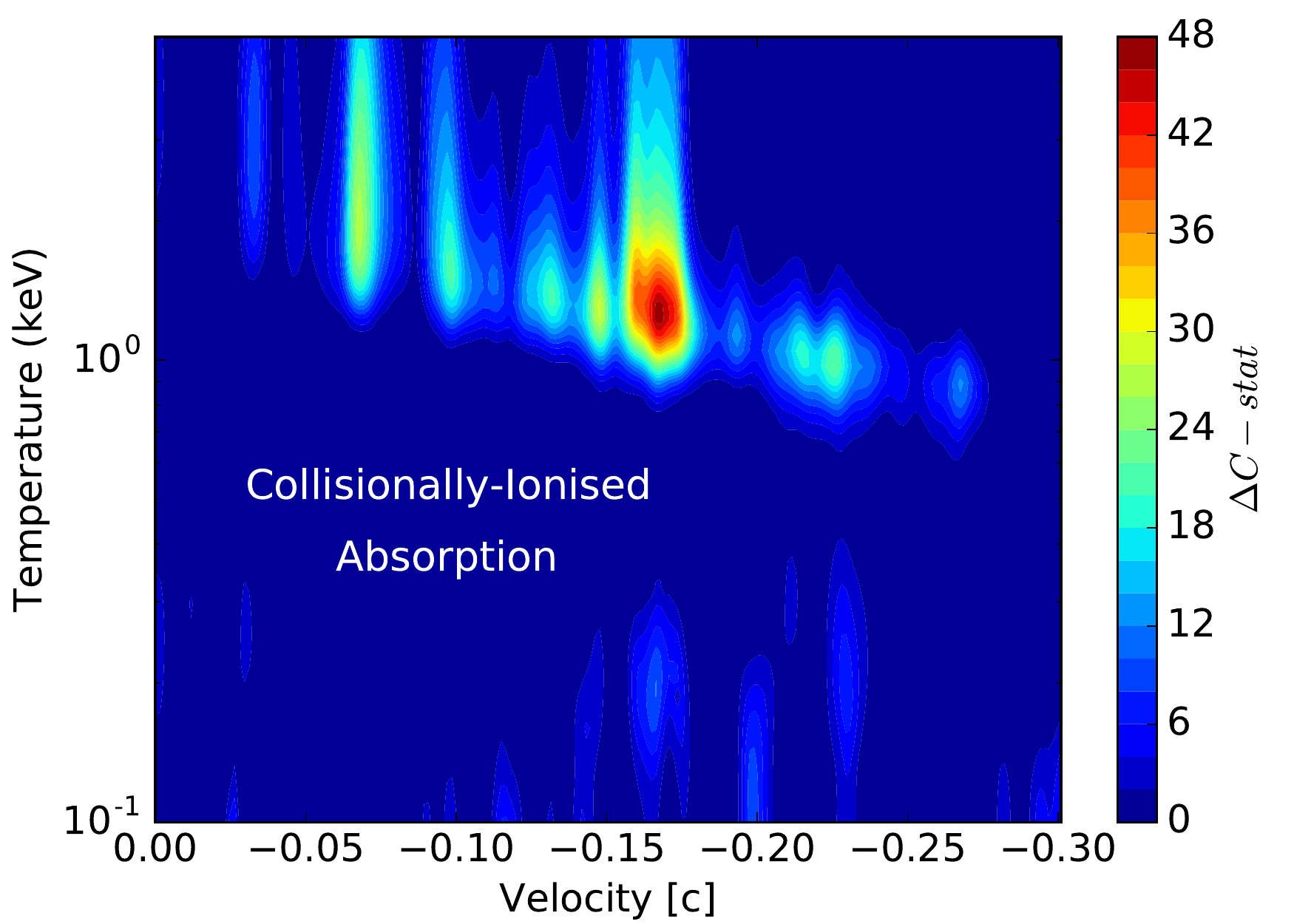}

   \vspace{-0.1cm}
  \caption{Multi-dimensional scans of collisional-ionisation emission (left panel) 
                \textcolor{black}{and absorption models}
                (right panel) for NGC 247 ULX-1 time-averaged EPIC+RGS spectra.
                The X-axis shows the line-of-sight velocity (negative means blueshift, {\it i.e.} motion towards 
                the observer). The color is coded according to the $\Delta C$ fit improvement to the spectral
                continuum model. }
   \label{Fig:emitting_gas_CIE}
   \vspace{-0.4cm}
\end{figure*}

We applied the automated routine scanning CIE models onto the NGC 247 ULX-1 
time-averaged RGS (0.33$-$2 keV) and EPIC MOS\,1,2-pn (1.77$-$10 keV) spectra. 
We adopted the three-blackbody continuum model shown in Sect.\,\ref{sec:baseline_continuum} 
and Table\,\ref{table:continuum_timeavg} (see also black line in Fig.\,\ref{Fig:Plot_XMM_wind_bestfits}). 
%%%The adopted CIE parameter search space was identical to that used for SS 433.
\textcolor{black}{The results are shown in Fig.\,\ref{Fig:emitting_gas_CIE} (left panel)}. 
The best-fit corresponds to a large improvement with respect to the 
continuum model ($\Delta C=82$, for 4 additional degrees of freedom) and
was achieved for a CIE temperature of 0.9 keV and a small blueshift of around 
$6500$ km/s ($\sim0.022c$).
These results were driven by the strong lines that can be seen in the RGS spectrum 
$\lesssim1$ keV in Fig.\,\ref{Fig:Plot_line_search} (bottom two panels).

\textcolor{black}{We have checked our method by testing an identical CIE scan 
on the well-known Galactic X-ray source SS 433.
This source is not ultraluminous in the X-ray band due to obscuration
of the accretion disk from local circumstellar gas, 
but exhibits a persistent and bright ($10^{40}$ erg/s)
radio jet, which is powering a luminous optical super-bubble \citep{Brinkmann1996,Fabrika2004,Medvedev2020}.
SS 433 is therefore considered to be viewed edge-on. 
Were it observed face-on, it would likely appear as a ULX 
(\citealt{Begelman2006,Poutanen2007,Middleton2018}).
SS 433 also shows a relativistic jet in the form of blueshifted lines from multi-temperature plasma
in collisional-ionisation equilibrium \citep{Marshall2002}.
The results obtained for SS 433 are very similar to NGC 247 ULX-1
and in agreement with \citet{Marshall2013} and \citet{Medvedev2018},
which supports our method (see Appendix \ref{sec:appendix_ss433}).}
%%%Moreover, \citet{Kosec2018a} showed that the strong emission features observed in the 
%%%RGS time-averaged spectrum of NGC 5204 ULX-1 are best described by blueshifted CIE plasma
%%%and that in some ULXs the emission lines might be from the jet.

\subsubsection{Collisionally-ionised absorbing gas}
\label{sec:appendix_hot}

\textcolor{black}{It is uncommon to adopt absorption models of gas 
in collisional-ionisation equilibrium in accreting objects as} 1) it is difficult to distinguish between 
photoionisation and collisional ionisation on the sole basis of the dominant resonant absorption lines
and 2) we hardly expect any jet to absorb the X-ray source continuum along our line of sight.
However, we cannot exclude that shocks are produced by interaction between the ULX wind and
the stellar companion or the surrounding bubble (or within the wind itself). 
Therefore, we also performed a model scan with the {\it hot} model in {\scriptsize{SPEX}},
which works just like {\it cie} but assumes \textcolor{black}{absorbing} gas.

\textcolor{black}{In Fig. \ref{Fig:emitting_gas_CIE} (right panel) we show the results 
using the {\it hot} model over the same $kT$ range used for the emitting gas,
adopting a velocity dispersion of 500 km/s and line-of-sight velocities, $v_{\rm LOS}$, 
ranging between $-0.3c$ and zero, {\it i.e.} only Doppler blueshifts
or outflows rather than inflows.
The best-fit solution is obtained for a $-0.17c$ blueshifted with a remarkable $\Delta C=48$
as suggested by the detection of several negative Gaussians blueshifted by similar values
in Fig.\,\ref{Fig:Plot_line_search}.}

\subsection{Photoionisation wind modelling}
\label{sec:photoionisation_modelling}

The emission and absorption lines can be produced by winds
rather than by jets as expected in the case of super-Eddington accretion discs and, therefore, in ULXs.
Accurate photoionisation models require knowledge of the radiation field,
{\it i.e.} the SED from optical to hard X-ray energies.

\subsubsection{SED and \textcolor{black}{photoionisation balance}}
\label{sec:SED}

Following \citet{Pinto2020a,Pinto2020b}, we built the time-averaged SED of 
NGC 247 ULX-1 using data from the XMM-\textit{Newton} campaign and archival HST 
observations (as taken from \citealt{Feng2016}). 
For issues regarding the non simultaneity 
of HST and XMM observations, see Appendix \ref{sec:appendix_SED}.
For the X-ray band ($0.3-10$ keV or $\sim10^{16-18}$ Hz) 
we used the best-fit three-blackbody continuum model, 
estimated in Sect.\,\ref{sec:baseline_continuum}, 
Fig.\,\ref{Fig:NGC247_RGS_EPIC_all_YLOG} and Table\,\ref{table:continuum_timeavg}.
As shown in Sect.\,\ref{sec:om_reduction}, the OM filters were not sensitive enough
to detect the optical counterpart, but their
flux upper limits in the optical and UV energy bands are however comparable to
the HST measurements (see Fig.\,\ref{Fig:Plot_SED_Balance}
top panel). We therefore modelled the optical/UV portion of the SED
with the two-blackbody model of \citet{Feng2016}, which together with the three-blackbody
X-ray model formed our five-blackbody SED model.
%%%For more detail on how the SED was constructed, see Appendix\,\ref{sec:appendix_SED}. 

\textcolor{black}{We can describe the photoionisation equilibrium with the ionisation parameter, $\xi$, 
defined as $\xi = L_{\rm ion} / ({n_{\rm H} \, R^2})$ (see, {\it e.g.}, \citealt{Tarter1969}),
where $L_{ion}$ is the ionising luminosity (measured between 
13.6 eV and 13.6 keV), $n_{\rm H}$ the hydrogen volume density and 
$R$ the distance from the ionising source. 
The ionisation balance was calculating with the {\scriptsize{SPEX}}
\textit{pion} model, which calculates the transmission and the emission of a thin slab of photoionised 
gas, self-consistently.}
%%%We also used the \textit{xabsinput} tool in {\scriptsize{SPEX}}, which performs
%%%the same calculation and provides in output the ionisation balance in a format
%%%useful for alternative photoionisation absorption models (see Sect. \ref{sec:xabs_gas}).
%%%The results obtained with \textit{pion} and \textit{xabsinput} are consistent.
%%%The ionisation balance is shown in Fig.\,\ref{Fig:Plot_SED_Balance} 
%%%(middle panel) for both the full five- and simple two-blackbody models. 
%%%There are small changes in the ionisation balance at intermediate $\xi$ values 
%%%between the full SED (five-\textit{bb} model) and the simple two-\textit{bb} X-ray model,
%%%but none of the best-fitting plasma solutions are affected.

\textcolor{black}{Following \citet{Pinto2020b}, we also computed the stability 
(or $S$) curve, which is the relationship between the temperature (or the 
ionisation parameter) and the ratio between the radiation and the thermal pressure,
which can be expressed as $\Xi = F / (n_{\rm H} c kT) = 19222 \, \xi / T$
\citep{Krolik1981}.
The stability curve is shown in 
Fig.\,\ref{Fig:Plot_SED_Balance} (bottom panel). 
The branches of the $S$ curve with a negative gradient are characterised by 
thermally unstable gas.}
In this work, we assumed that the wind is seeing the same SED that we observe
and, therefore, adopted the five-blackbody model SED and ionisation balance.
Systematic effects from the SED choice are discussed in Sect.\,\ref{sec:discussion} 
and Appendix\,\ref{sec:appendix_SED}.

\subsubsection{Photoionised emitting gas}
\label{sec:pie_gas}

Once the SED and the ionisation balance were computed, we scanned the time-average
EPIC+RGS spectra with the {\scriptsize{SPEX}} \textit{pion} model with the same multi-dimensional 
routine used for the \textit{cie} model in Sect.\,\ref{sec:cie_ngc247}, and a similar parameter space.
We adopted a logarithmic grid of ionisation parameters (log\,$\xi$ [erg/s cm] between 0 and 6 
with 0.1 steps).
The only free parameter for the \textit{pion} is the column density, $N_{\rm H}$. 

Unlike NGC 1313 ULX-1, the RGS spectrum of NGC 247 ULX-1 does not exhibit well resolved  
 emission line triplets. This is perhaps due to the longer integration time required and 
the variability of the line centroid (see Fig. \ref{Fig:Plot_EPIC_spectra_res}), 
which could wash out the triplets when stacking all the spectra.
Additionally, the crucial {\ovii} complex is affected by the background noise.
The lack of He-like triplets means that the volume density and the luminosity of the photoionised
gas are degenerate. Fitting both parameters results in poor constraints and much higher 
computation time.
We therefore chose not to fit the volume density and adopted $n_{\rm H} = 10^{10}$ cm$^{-3}$,
which is a lower limit found for NGC 1313 ULX-1 \citep{Pinto2020b}.
This \textcolor{black}{would only slightly affect the overall flux and column density of the \textit{pion} 
component.}

The \textit{pion} covering fraction is set to zero
(i.e \textit{pion} only produces emission lines)
and the solid angle $\Omega=4\pi$.
Fitting additional parameters such as $\Omega$ might provide even better fits but 
would significantly increase the computing time,
without altering the velocity and ionisation parameters.

\begin{figure*}
 \centering
  \includegraphics[width=0.99\columnwidth, angle=0]{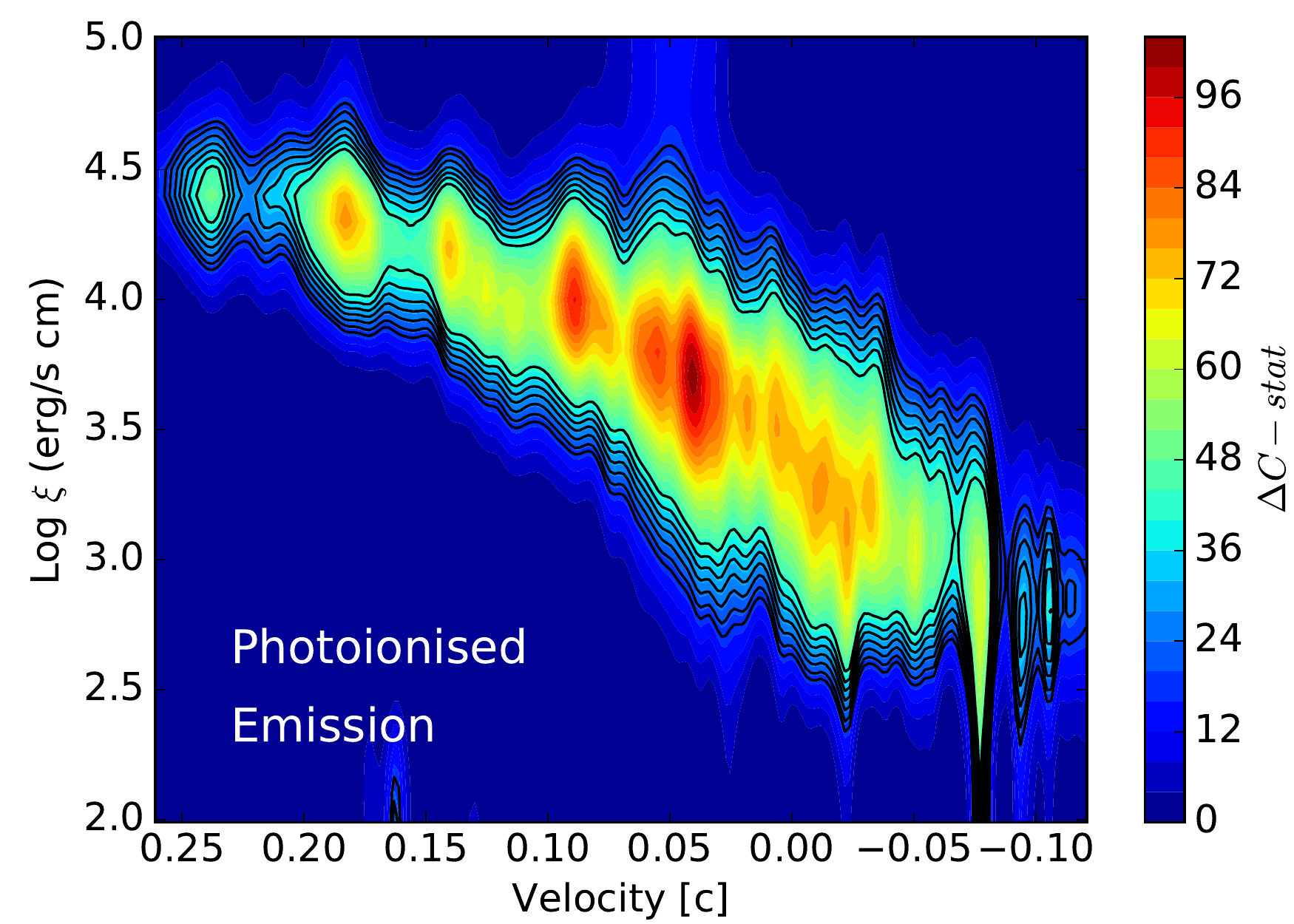}
  \includegraphics[width=0.99\columnwidth, angle=0]{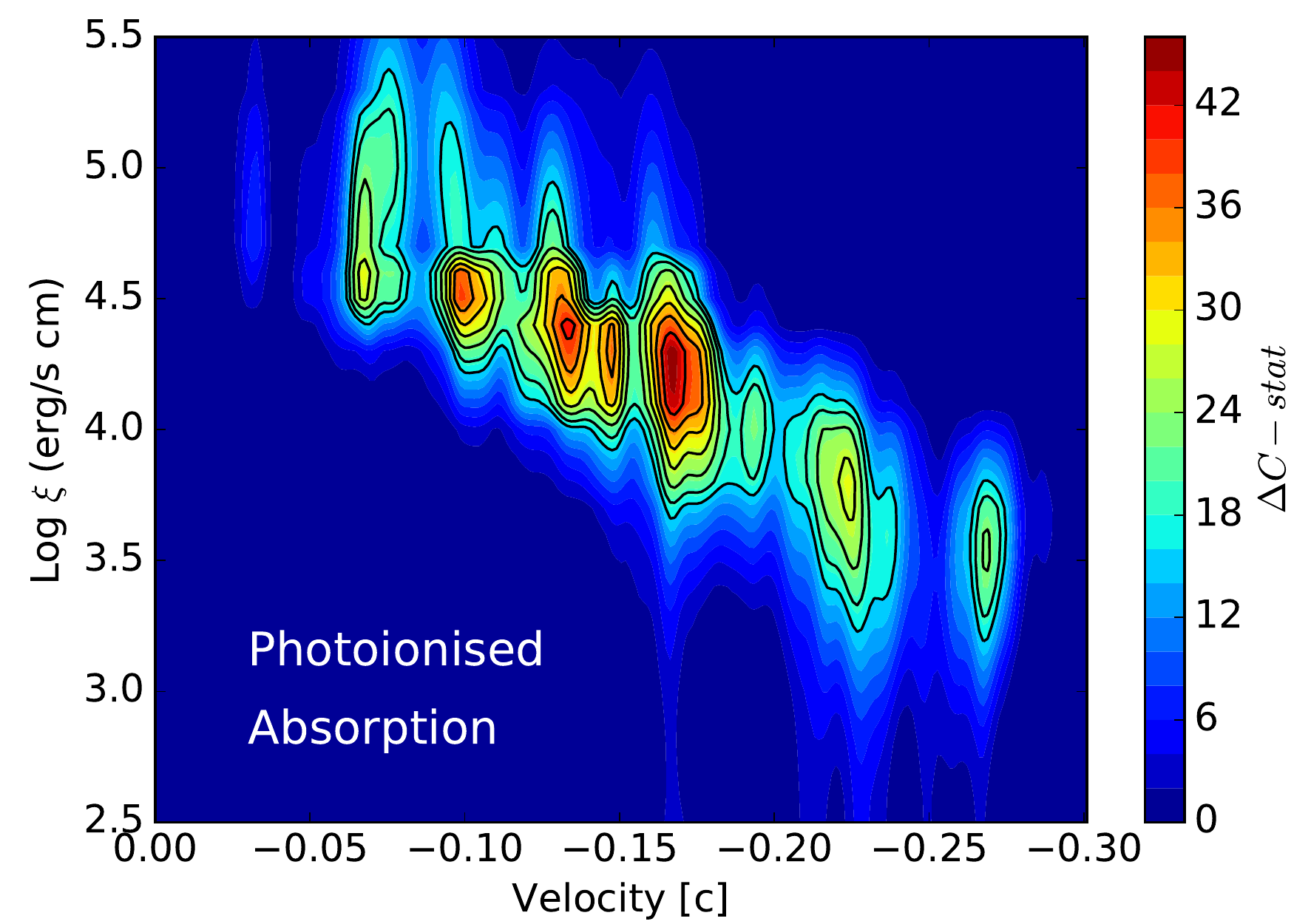}
  \vspace{-0.1cm}
   \caption{\textcolor{black}{Scans of photoionisation emission (left) and absorption (right) 
                models for the time-averaged EPIC and RGS spectra.
                \textcolor{black}{Labels are same as in Fig. \ref{Fig:emitting_gas_CIE}.}
                The black contours refer to the (2.0, 2.5, ...
                4.5, 5.0)\,$\sigma$ confidence levels estimated by
                Monte Carlo simulations.}}
%   Further detail on the properties of the absorbers and the model scan
%   can be found in Fig.\,\ref{Fig:photoionised_gas_part_2} and Sect.\,\ref{sec:absorbing_gas}.}
   \label{Fig:photoionised_gas}
  \vspace{-0.4cm}
\end{figure*}

In Fig.\,\ref{Fig:photoionised_gas} (left panel) we show the results obtained 
using a \textit{pion} line width of 1000 km/s. 
The confidence level (CL) is expressed in $\sigma$, which 
is constrained using Monte Carlo simulations
(see Sect.\,\ref{sec:mc_simulations}). 
A peak ($\Delta C = 77$) corresponding to a solution of blueshift emission is seen around 
$0.02-0.03c$ in agreement with the CIE model scans (see Sect.\,\ref{sec:cie_modelling}
and Fig.\,\ref{Fig:emitting_gas_CIE}). However, the different ionisation balance and types of emission 
lines in the photoionisation equilibrium detected another, stronger ($\Delta C = 102$), 
peak corresponding to a redshift of $\sim+0.05c$. 

\subsubsection{Photoionised absorbing gas}
\label{sec:xabs_gas}

In principle, we could just use \textit{pion} for both emission and absorption. 
However, this model re-calculates the ionisation balance 
at every iteration and therefore is computationally expensive. 
Therefore, for the absorbing gas we chose
to use the faster \textit{xabs} model, which is optimized for absorption and adopts
the pre-calculated ionisation balance (see Sect.\,\ref{sec:SED}).

The \textit{xabs} model shares several parameters with \textit{pion} except the
opening angle of the line emission which is zero since no emission is present in 
this model. We adopted a covering fraction equal to unity
in order to avoid degeneracy and reduce the computing
time. \textcolor{black}{We calculated the grid of photoionised \textit{xabs} models in the same way 
as the \textit{pion} models, but assuming line-of-sight velocities, $v_{\rm LOS}$, 
ranging between $-0.3c$ and zero, {\it i.e.} only Doppler blueshifts
as for the CIE {\it hot} absorption models in Sect. \ref{sec:appendix_hot}.}

In Fig.\,\ref{Fig:photoionised_gas} (right panel) we show the probability distributions 
from scans of the RGS and EPIC spectra with 
$v_{\rm RMS} = 1000$ km/s. As expected, our code confirmed
the $v=-0.17c$ solution ($\Delta C = 46$). 

\subsection{Final fits with physical plasma models}
\label{sec:model_comparison}

\begin{figure}
 \centering
 \includegraphics[width=\columnwidth, angle=0]{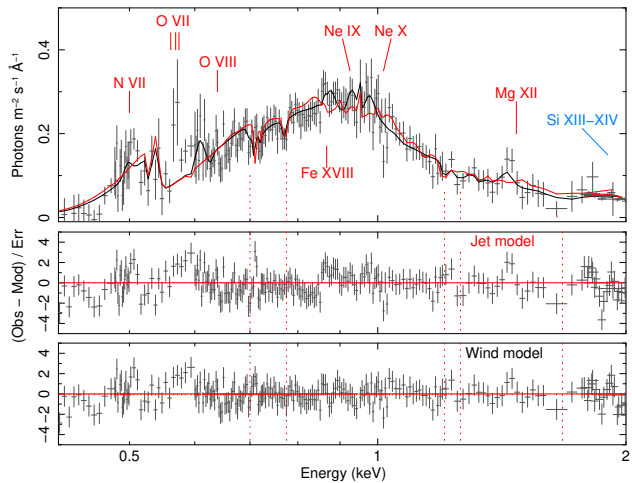}
  \vspace{-0.4cm}
   \caption{Time-averaged XMM-\textit{Newton} RGS (0.33$-$2 keV) and EPIC-pn (1.77$-$10 keV)
                 spectra. Overlaid are two alternative models (jet - CIE, red line
                 and the wind - PIE, black line).
                 The spectra were regrouped
                 and the plot zoomed onto the RGS data
                 and the 0.4--2 keV energy band for displaying purposes.}
   \label{Fig:Plot_XMM_wind_vs_jet_zoom}
  \vspace{-0.4cm}
\end{figure}

In order to check the inter-dependence of the emitting and absorbing plasma components
and to test for any variations in the values of their parameters we performed two more fits
of the RGS and EPIC data (with EPIC still excluded between 0.33--1.77 keV) adding onto
the 3-blackbody continuum two alternative plasma models. The first one was a wind model 
that used the {\it pion} in emission and the {\it xabs} in absorption. 
The second model was an approximation of jets and shocks in the 
form of {\it cie} component in emission and {\it hot} component in absorption.
The results for the two models are shown in Table \ref{table:plasma_components}
and Fig. \ref{Fig:Plot_XMM_wind_vs_jet_zoom} 
\textcolor{black}{(zoomed onto the RGS)}.
The absorption components provided very similar results, especially for the column densities
and the line-of-sight velocities. As previously noted, the emission components 
show some differences. The results are discussed in Sect. \ref{sec:discussion}.

\textcolor{black}{The temperatures of the three blackbody components were always fixed to the 
EPIC 0.3--10 keV fits (Table \ref{table:continuum_timeavg}) each time we ignored EPIC data 
between 0.33--1.77 keV, resulting in parameters consistent with the continuum modelling.}

%--------------------------------------------TABLE START---------------------------------------------

\begin{table}
\caption{\textcolor{black}{NGC 247 ULX-1: alternative plasma models.}}  
\label{table:plasma_components}             % is used to refer this table in the text
\renewcommand{\arraystretch}{1.}      % z = sqrt ( (1. + v/c) / (1. - v/c) ) -1
 \small\addtolength{\tabcolsep}{-0.5pt}
 
\scalebox{1}{%
\hskip-0.5cm\begin{tabular}{@{}l l l l l l l l}     
\hline  
Model 1                        & Parameter                      &  Emission                        &  Absorption      \\  
\hline                                                                  
\multirow{4}{*}{PIE}      &$L_{\rm X}(E)$ , $N_{\rm H}(A)$& $1.4\pm 0.2$      & $2.8 \pm 0.1$     \\
\multirow{5}{*}{(wind)}  &   $\log \xi$  (erg/s cm)   & $3.7 \pm 0.1$                 & $4.3 \pm 0.1$     \\
                                     &   $v_{\rm LOS}\,(c)$       & $+0.042\pm0.004$          & $-0.166\pm0.002$ \\
                                     &   $v_{\rm RMS}$  (km/s)    & $3000_{-800}^{+2700}$ & $400\pm 200$   \\
                                     &   $\Delta C^a(^e)$          & 104(97)                          & 46(32)                 \\
                                     &   $\sigma^a(^e)$            & $>5$($>5$)                    & $>5$($>4$)         \\
\hline                                                                  
%%%\hline                                                                  
Model 2                        & Parameter                      &  Emission             &  Absorption            \\  
\hline                                                                  
\multirow{4}{*}{CIE}      &$L_{\rm X}(E)$ , $N_{\rm H}(A)$&  $1.4 \pm 0.2$   &  $2.5 \pm 0.1$ \\
\multirow{5}{*}{(jet)}      &  $kT$  (keV)                   &  $0.9 \pm 0.1$      &  $1.3 \pm 0.1$     \\
                                     &   $v_{\rm LOS}$ ($c$)    & $-0.022\pm0.002$  & $-0.168\pm0.003$ \\
                                     &   $v_{\rm RMS}$  (km/s)    & $3000_{-1000}^{+3000}$ &  $1600\pm850$  \\
                                     &   $\Delta C^a(^e)$          & 83(51)                   & 48(32)                   \\
                                     &   $\sigma^a(^e)$             & $>5$($>5$)          & $>5$($>4$)          \\
\hline                                                                                         
%%%\hline                                                                                                                
\end{tabular}}

\vspace{0.1cm}

\textcolor{black}{Main properties of the plasma components for two alternative models. 
Model 1 or PIE/wind: photoionised emission ({\it pion}) and absorption ({\it xabs}, see 
Fig.\,\ref{Fig:photoionised_gas} and Fig.\,\ref{Fig:Plot_XMM_wind_vs_jet_zoom}).
Model 2 or CIE/jet: collisionally-ionised emission ({\it cie}) and absorption ({\it hot}, 
see Fig.\,\ref{Fig:emitting_gas_CIE}).
The column densities, $N_{\rm H}$, are in 
$10^{22}\,{\rm cm}^{-2}$. The line-of-sight velocities, $v_{\rm LOS}$,
are in units of light speed $c$; 
the velocity dispersion, $v_{\rm RMS}$, is in km s$^{-1}$.
The 0.3--10\,keV luminosities of the emitting plasmas, $L_{\rm X}(E)$, are defined 
in units of $10^{38}$ erg/s.
%%%, and correspond to $n_{\rm e}n_{\rm H}V=(5.9\pm0.6)\times10^{60}\,{\rm cm}^{-3}$
%%%for the {\it cie} and $N_{\rm H}=(1.0\pm0.1)\times10^{22}\,{\rm cm}^{-2}$ 
%%%for the {\it pion}, respectively.
The $\Delta C^a(^e)$ refer to the $\Delta C$-statistics of each component computed 
when the component is the only one in model (a) or when the other one is included (e).
The same applies to the detection significances, $\sigma^a(^e)$, evaluated
with Monte Carlo simulations. }
   \vspace{-0.2cm}
\end{table}

%--------------------------------------------TABLE END---------------------------------------------

By alternatively excluding one particular plasma component from the spectral model
we estimated the relative contribution to the spectral fit and the minimum $\Delta C$-stat
improvement of each component for both the wind and jet model (see $\Delta C^e$ values in 
Table \ref{table:plasma_components}). Following \citet{Pinto2020b}, we compared the minimum 
$\Delta C$-stat values with the Montecarlo simulations to obtain an approximate estimate 
of the minimum significance of each wind or jet component (parameter $\sigma^e$),
which is highest for the emission phases.

Finally, to further test the strength of our results we performed a fit of the RGS and EPIC data
including the whole EPIC energy band but fixing the plasma models to the best-fit results
of the hybrid RGS (0.33-2 keV) plus EPIC (1.77-10 keV) fits. In Sect. \ref{sec:xabs_gas},
we noted that the inclusion of the photoionised {\it xabs} absorption component significantly
decreased the C-stat from 4488 (of the simple 3-blackbody model) to 3033.
The addition of the photoionised {\it pion} emission component further lowered the C-stat
to 2431 (with a $\chi^2=1830$ and a total of 1132 degrees of freedom).
We performed the same fit by using instead the RGS jet model with the {\it hot} 
collisionally-ionised component fixing the parameters to those in 
Table \ref{table:plasma_components}. This decreased the C-stat from 4488 to 3043, 
similarly to the photoionised absorber. The addition of a {\it cie} emission component
(with fixed plasma parameters) implied a final C-stat = 2542, which is slightly worse
than {\it pion} due to some positive residuals left around 0.9 keV that can also be seen
in the RGS spectral modelling in Fig. \ref{Fig:Plot_XMM_wind_vs_jet_zoom}. 

\textcolor{black}{We notice that all the spectral fits shown here are not formally acceptable
(see Table \ref{table:model_comparisons}), although the winds components provide 
significant improvements. 
One reason is the variability of the features, both of their 
centroids and relative strength (see Fig. \ref{Fig:Plot_EPIC_spectra} and 
\ref{Fig:Plot_EPIC_spectra_res}).
This means that more complex models would be required to correctly fit the lines.
On the other hand, the winds are likely multiphase as shown by the low-temperature 
{\ovii} clearly missed by our single phase model (see Fig. \ref{Fig:Plot_XMM_wind_vs_jet_zoom}).
This was already shown in \citet{Pinto2020b} and occurs in
SS 433 too (see Appendix \ref{sec:appendix_ss433}).
Finally, some bad cross-calibration below 0.6 keV between the EPIC and RGS cameras further 
prevent us from obtaining fully acceptable fits (see Fig. \ref{Fig:NGC247_RGS_EPIC_all_YLOG}).}

\begin{table}
\caption{\textcolor{black}{NGC 247 ULX-1: continuum and plasma models}}  
\label{table:model_comparisons}     
\renewcommand{\arraystretch}{1.1}
 \small\addtolength{\tabcolsep}{-2pt}
 
\scalebox{0.9}{%
\hskip-0.5cm\begin{tabular}{@{} l l l l}     
\hline
Model                                              & RGS+EPIC\,($>1.77$\,keV) &  RGS+EPIC\,(full band)      \\  
\hline               
$3\,bb$                                            & 2019/994                              & 4488/1132                          \\
$3\,bb*hot$                                      & 1971/990                              & 3043/1132  gas par fixed   \\
\multirow{2}{*}{$3\,bb*hot+cie$}      & \multirow{2}{*}{1895/886}     & 2542/1132  gas par fixed   \\
                                                        &                                              & 2233/1124                           \\
$3\,bb*xabs$                                   & 1973/990                              & 3033/1132  gas par fixed    \\
\multirow{2}{*}{$3\,bb*xabs+pion$} & \multirow{2}{*}{1877/886}     & 2431/1132  gas par fixed    \\
                                                        &                                              & 2191/1124                          \\
\hline                                                                                                                                                                                           
\end{tabular}}

\textcolor{black}{C-stat/d.o.f. values for spectral continuum and plasma models. `gas par fixed' means that for 
those fits the parameters of both emission and absorption components were fixed to the best-fit
values obtained excluding EPIC data below 1.77 keV.}
\vspace{-0.2cm}
\end{table}

\section{Discussion}
\label{sec:discussion}

\textcolor{black}{It is still unclear how does the wind vary with the accretion rate and whether it has
a major role in shaping ULX spectra.
\citet{Pinto2020a,Pinto2020b} showed that the wind evolves with the changes in the
continuum from the fainter, harder states to the brighter states, which implies a tight
relationship between the source's spectral continuum and wind appearance 
as observed by comparing winds in different ULXs.}

Among the ULXs, the supersoft ultraluminous X-ray (SSUL) sources
are particularly fascinating objects. The fact that such sources reach very high luminosities 
(several $10^{39}$ erg/s) but always exhibiting very soft ($kT\sim0.1$ keV) spectra
indicates that they are being observed at moderate-to-high inclination angles as 
also suggested by the presence of dips in their lightcurves (see, {\it e.g.}, \citealt{Feng2016}).
In fact, \citet{Urquhart2016} modelled the soft X-ray residuals and the $\sim1$ keV drop 
found in several CCD spectra of ULSs with a model of thermal emission and an absorption
edge, which they interpreted as a result of absorption and photon reprocessing by an 
optically-thick wind which obscures the innermost regions where most hard X-rays are produced.

%%%Motivated by the recent detection of relativistic winds in several ULXs and the presence 
%%%of very strong residuals in the CCD spectra of NGC 247 ULX-1, 
%%%we requested and were awarded 7 orbits with XMM-\textit{Newton} during 
%%%AO18 (PI: Pinto) with the primary aims of 1) detecting a wind through the RGS emission
%%%and/or absorption lines, 2) studying the evolution of the continuum and the residuals in great 
%%%detail with time- and flux-resolved spectroscopy, and 3) searching for preferred timescales
%%%in the Fourier spectrum of the source and possible links with the dipping activity.
%%%This was the first paper in a series focusing on NGC 247 ULX-1, and presented the search
%%%for a wind taking advantage of the high-resolution X-ray spectrometers.
%%%Detailed broadband spectroscopy and analysis of continuum evolution throughout the several
%%%epochs is performed in D'A\`{i} et al. (in prep), while the Fourier analysis is done by 
%%%\citet{Alston2021}. Further work will focus on a flux-resolved high-resolution X-ray 
%%%spectroscopy and the study of the source populations in the NGC 247 galaxy.

\subsection{Time evolution of the 1 keV residuals}
\label{sec:discussion_1keV}

The stacked XMM-\textit{Newton} lightcurve (see Fig. \ref{Fig:Plot_XMM_lc}) shows
different pattern of source variability such as a long-term overall change in the average flux 
on daily time scales followed by abrupt drops in the flux where the source becomes softer
(the dips) on timescales between 100s and a few hours.
The dipping activity seems also to enhance during observations with higher flux peaks. 
The higher flux might be associated with a higher local accretion rate, which then 
would increase the radiative force and launch optically-thick wind cloudlets in the line of sight,
thereby obscuring the innermost regions of the disc responsible for the hard X-ray emission
(as suggested by \citealt{Urquhart2016}).
More insights on the nature of the dips will be provided by \citet{Alston2021}.
This work shows that the dips in the higher flux observations preferentially occur on $5$ and $10$\,ks timescales, which suggests that \textcolor{black}{they are caused from obscuration at $\sim 10^{4-5} R_{\rm G}$, 
where $R_{\rm G}$ is the Gravitational radius 
(if the timescales are associated with keplerian motion around a NS or a stellar-mass BH)}. 
Such a range is comparable to the distance
that the $0.17c$ wind would travel on a time scale of 1\,ks, suggesting a possible connection
between them.
%%%\textcolor{red}{\textbf{CP: Will shall we highlight the coolest result you've found? 
%%%If yes can you send me a sentence summarising it in a discussion-like shape?}}

The high-quality EPIC spectra of the individual observations show a remarkable flux 
variability
in the features around 1 keV (see Fig. \ref{Fig:Plot_EPIC_spectra}). 
In order to quantify such variability, we modelled the two strongest absorption 
features around 0.7 and 1.2 keV, and the dominant emission-like feature at 1 keV with three Gaussian
lines for the EPIC (MOS and pn) spectral of the individual observations. 
All lines show a distinct pattern with their energy centroids significantly blueshifted during the 
brightest observations (which also exhibit most dips and the highest variability, 
see Fig. \ref{Fig:Plot_EPIC_spectra_res}).
Interestingly, the fluxes of the high-energy lines (1 and 1.2 keV) significantly decrease
during the dipping observations while the 0.7 keV line 
\textcolor{black}{seems to strengthen (see Table \ref{table:continuum_allabs}).}
This would either suggest a different location of the three lines, with the 0.7 keV line
coming from the outer and less obscured regions, or a change in the ionisation state 
of the absorber during \textcolor{black}{the high-flux periods}.
This is similar to what was observed in NGC 1313 ULX-1 (\citealt{Pinto2020b}, \citealt{Middleton2015b}).
A detailed study of the broadband spectra and residuals evolution
will be shown by D'A\`{i} et al. (in prep).
%%%\textcolor{red}{\textbf{CP: Anto shall we highlight the coolest result you've found? 
%%%If yes can you send me a sentence summarising it in a discussion-like shape?}}
The fact that the location and strength of the residuals vary on hourly timescales 
with the source flux provides strong evidence in support 
for a disc wind rather than emission from the local ULX bubble or the galactic ISM.

\subsection{A wind or a jet?}
\label{sec:discussion_wind}

\subsubsection{Emission lines}
\label{sec:discussion_wind_emission}

\textcolor{black}{The time-average stacked RGS spectrum 
showed} strong emission residuals near the transition energies of several 
ions such as {\oviiviii}, {\neixx} and {\fexviii} 
(see Fig.\,\ref{Fig:Plot_XMM_wind_bestfits_zoom} and \ref{Fig:Plot_line_search}).
The agreement between RGS and EPIC is corroborated 
by applying the wind model constrained using only RGS in the $0.33-1.77$ keV
band to the whole EPIC MOS and pn time-average spectra (see 
Table \ref{table:model_comparisons}).
Unfortunately, the He-like emission triples of {\it e.g.} {\ovii} and {\neix} are not well resolved
likely due to the stacking of RGS spectra from different 
observations that clearly showed some variability in the line centroid as discussed above.
This limited our capabilities of distinguishing between collisional and photoionisation,
but the use of full plasma models provided some constraints.

By performing automated searches of plasma models in a large parameter space, we built
probability contours for both collisionally-ionised and photoionised plasma emission models.
The properties of the line-emitting gas are very similar to those of the Galactic super-Eddington
source SS 433 with a low velocity along the line of sight and a mild 1 keV temperature which is
expected by the strong Ne K and Fe L emission around 1 keV (see Fig. \ref{Fig:emitting_gas_CIE}).
It is well established that the emission lines of SS 433 are from the jet with the low velocity 
indicating that the precessing jet was at very high angle, close to 90 degrees,
in the analysed observation. If the lines of NGC 247 ULX-1 were also from a variable jet,
the observed low velocity would suggest that it is being viewed at high angle in agreement
with the presence of dips.

The photoionisation emission models ({\it pion} component in {\scriptsize{SPEX}}, see Fig. \ref{Fig:photoionised_gas}) however provided a significantly
higher improvement to the spectral fits and a better description of the emission lines
(see Table \ref{table:plasma_components} and Fig. \ref{Fig:Plot_XMM_wind_vs_jet_zoom}).
This together with the evolution of the lines with the source continuum would favour 
photoionisation equilibrium \textcolor{black}{similarly to} the emission lines in NGC 1313 ULX-1
(\citealt{Pinto2020b}).

Regardless of the adopted equilibrium state, the luminosity of the line-emission component 
is remarkably high ($L_{\,0.3-10\,\rm keV}>10^{38}$\,erg/s), similarly to NGC 1313 ULX-1, 
NGC 5408 ULX-1 \citep{Pinto2016nature}, NGC 55 ULX-1 \citep{Pinto2017}, NGC 5204 ULX-1
\citep{Kosec2018a} and other ULXs ({\it e.g.}, \citealt{Wang2019}). This is about 2-3 orders of 
magnitude higher than the emission lines in SS 433 and those producing the winds
in classical supergiant X-ray binaries (sub-Eddington neutron stars accreting 
from supergiant OB stars, {\it e.g.} \citealt{ElMellah2017}) or 
the lines from accretion disc coronae of low-mass X-ray binaries 
(see, {\it e.g.}, \citealt{Psaradaki2018}).
The luminosity of $1.4\times10^{38}$\,erg/s is instead comparable to the extended X-ray
emission recently found around the extremely bright pulsating NGC 5907 ULX-1
\citep{Belfiore2020}, which suggests that the wind might be energetic enough to mechanically
drive the $\sim100$-pc super bubbles (see also \citealt{Pinto2020a}).

\textcolor{black}{Similarly to NGC 1313 ULX-1, the {\ovii} emission lines cannot be reproduced with the 
emission component responsible for the Fe L and Ne K emission 
(see Fig. \ref{Fig:Plot_XMM_wind_vs_jet_zoom}). A second component 
(either photo- or collisionally-ionised) with a low blueshift of 6000 km/s would be required.}

\subsubsection{Absorption lines}
\label{sec:discussion_wind_absorption}

In this work we also reported a highly significant detection of mildly-relativistic, ultrafast,
outflows. Both collisional and photoionisation (see Fig. \ref{Fig:emitting_gas_CIE}, 
\ref{Fig:photoionised_gas} and \ref{Fig:Plot_XMM_wind_vs_jet_zoom}) 
plasma models identified a high velocity outflow
($-0.17c$) in the range of the velocities found in other ULX winds.

The ionisation parameter is rather high (log $\xi = 4.3$) which is not surprising given the
soft SED adopted for this source (see Fig. \ref{Fig:Plot_SED_Balance}). If the wind at the
launch is seeing a different SED ({\it e.g.}, the hot innermost regions presumably obscured
along our line of sight) the overall ionisation balance might be significantly different.
This subject was extensively discussed in \citet{Pinto2020a} who found larger instability
branches in the $S$ curves of harder ULXs. Therefore, as a test, we performed an 
additional fit with the photoionised {\it pion {\rm +} xabs} wind model (as previously done 
in Sect. \ref{sec:model_comparison}) by adopting the ionisation balance calculated 
for the hard state of NGC 1313 ULX-1 in \citet{Pinto2020b} to estimate the systematic 
effects on the wind parameters.
The fit was statistically indistinguishable from the one performed with the ionisation balance
computed for NGC 247 ULX-1, with the exception of the ionisation parameters which, as expected,
turned out to be significantly lower by about $\Delta$\,log\,$\xi \sim 1$.

The absorption lines are generally weaker than the emission lines in the RGS spectrum 
of NGC 247 ULX-1 which could be due to the low source continuum (Kosec et al. submitted). 
This was also predicted by \citet{Pinto2017} as the lines are normally seen
on top of the continuum from the innermost regions which in this case is likely obscured.
 
Statistically we cannot distinguish photoionisation from collisional ionisation, but the former 
is favoured by the photoionised nature of the emitting plasma and the unusual detection of 
collisionally-ionised absorption in XRB winds.
 
\subsection{Accretion disc and wind physics}
\label{sec:discussion_system}

\begin{figure*}
 \centering
  \includegraphics[width=0.9\columnwidth, angle=0]{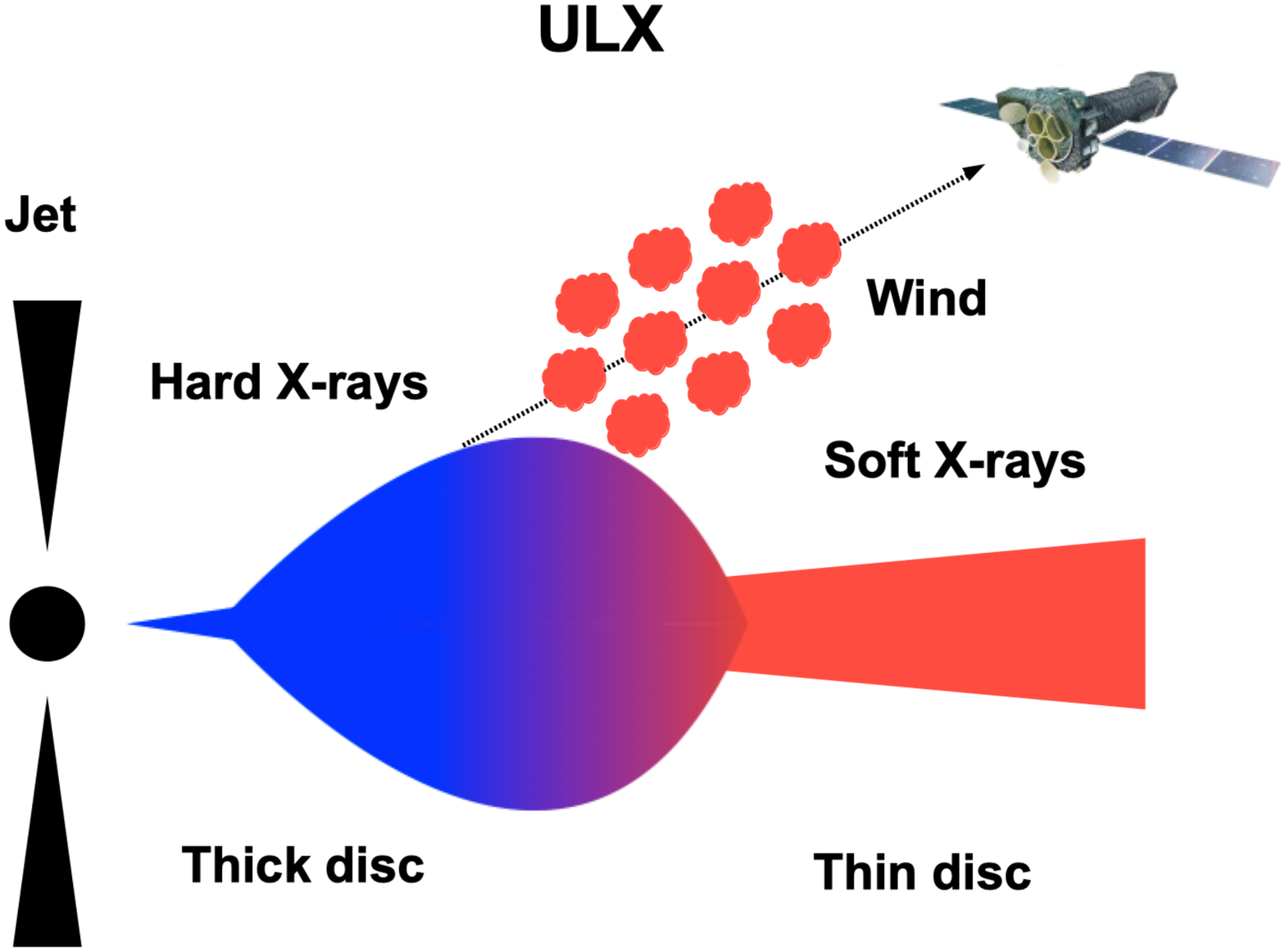}
  \hspace{0.5cm}
  \includegraphics[width=0.9\columnwidth, angle=0]{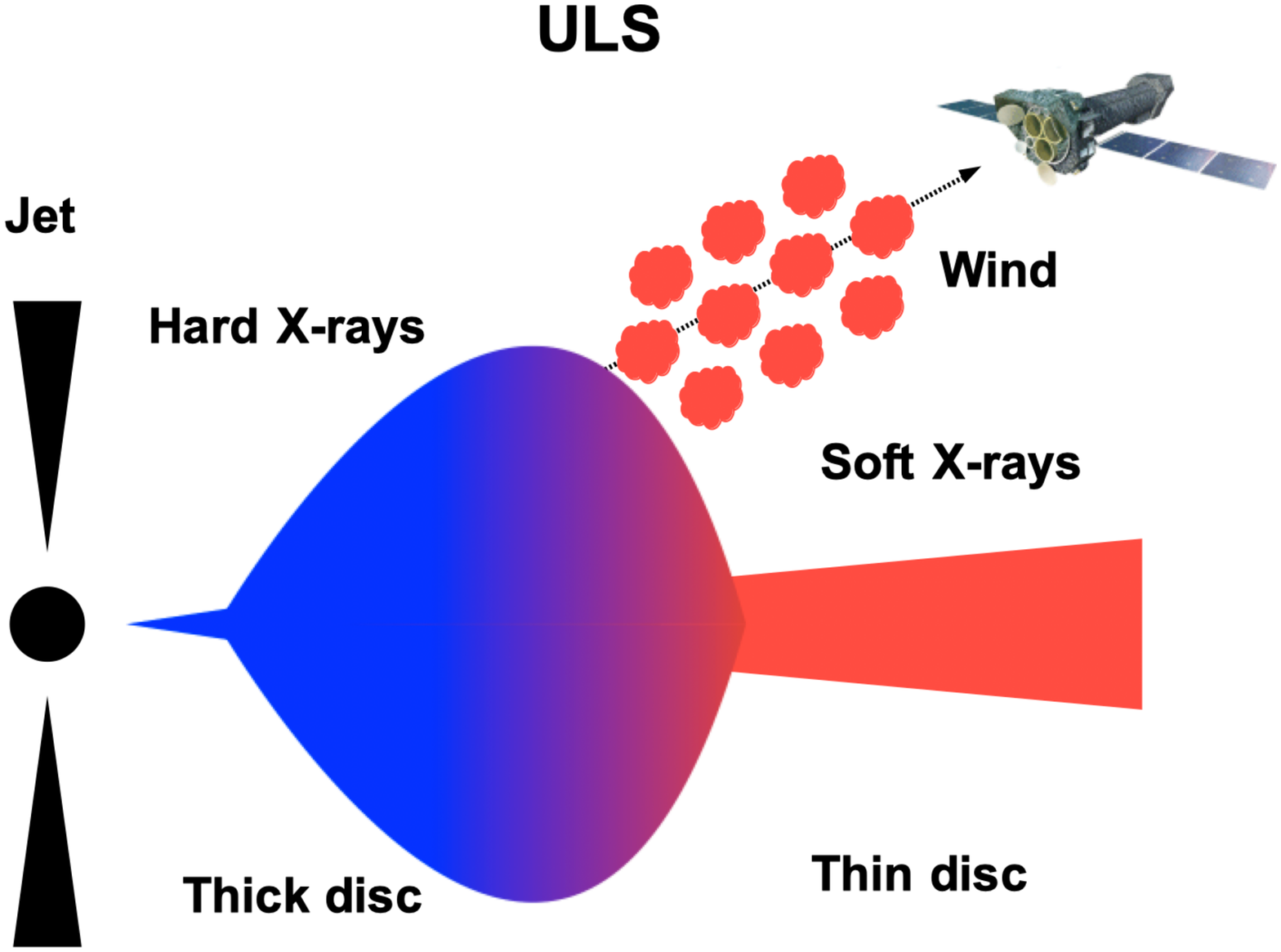}
  \vspace{-0.3cm}
   \caption{A possible scenario for the dips and ULX-ULS transitions in NGC 247 ULX-1.
                 The source is observed at a viewing angle that is high enough that 
                 the inner disc is already partly obscured by the wind (soft ULXs, left panel).
                 An increase of the accretion rate pushes up the scale-height of the disc and the optically-thick
                 base of the wind, causing an near-total obscuration of the inner regions and the source appears
                 as an ultraluminous supersoft source (ULS, right panel,
                see also \citealt{Pinto2017,Pinto2020b}, \citealt{Guo2019}). }
   \label{Fig:big_picture}
  \vspace{-0.4cm}
\end{figure*}

In the framework of super-Eddington accretion the luminosity scales with the logarithm of the
accretion rate in Eddington units ($\dot{m}=\dot{M} / \dot{M}_{\rm E}$) times the geometrical 
beaming of the funnel created by the height of the 
disc around the spherisation radius and by the wind itself (see Fig. \ref{Fig:big_picture}).
Following \citet{King2020}, the apparent luminosity can be expressed with 
\textcolor{black}{$L_{\rm app}=L/b=L_{\rm E}(1+\ln \dot{m})/b$, where $L$ is the intrinsic}
luminosity, $L_{\rm E}$ the luminosity in Eddington units, and $b=73/\dot{m}^2$ the geometrical
beaming. 

To estimate the bolometric luminosity of NGC 247 ULX-1 we integrated the broadband 
SED between 1\,eV and 10\,keV (or $2.4\times10^{14-18}$ Hz, see Fig. \ref{Fig:Plot_SED_Balance}) 
and obtained $9.4\times10^{39}$ erg/s. NGC 247 ULX-1 luminosity
could therefore be explained by assuming a black hole accreting above 10 times
the Eddington rate or a neutron star accreting above $\dot{m}=25$.
At $\dot{m}\sim10$ the spherisation radius, {\it i.e.} the base of the wind, would
be $R_{\rm sph}=27/4 \dot{m} R_{\rm G}=68R_{\rm G}$.
Interestingly, this is very close to the escape radius for a $-0.17c$ wind 
($R_{\rm e}=2GM/v^2=2R_{\rm G}c^2/v^2=73R_{\rm G}$), which would 
indeed suggest that we detected a wind launched from the spherisation radius
of a compact object above 10 $\dot{M}_{\rm E}$.

From Eq. (38) in \citet{Poutanen2007}, assuming $M_{\rm BH}=10M_{\odot}$ and $\dot{m}=10$, 
we estimated a temperature for the spherisation radius $T_{\rm sph}\sim0.3$ keV, 
which is comparable to the warm blackbody component in our fits 
(see Table \ref{table:continuum_timeavg}),
with the cooler ($\sim0.1$ keV) blackbody associated with the outer disc and, likely, the
wind photosphere as suggested by recent work (see, {\it e.g.}, \citealt{Qiu2021}, 
\citealt{Gurpide2021}).
We notice, however, that the source is being seen at high inclination with a substantial 
fraction of the hard X-ray photons obscured by the funnel. 
The intrinsic luminosity of NGC 247 ULX-1 might therefore be higher than the value 
estimated above with a higher
accretion rate, implying $T_{\rm sph}\sim0.1-0.2$ keV, closer to the cooler
blackbody component, and a slightly larger $R_{\rm sph}$.
It is also possible that the wind is launched with lower velocity at radii larger 
than $73R_{\rm G}$ and it gets accelerated by radiation pressure from the inner
accretion flow (see, {\it e.g.}, \citealt{Takeuchi2013}).

\textcolor{black}{Similar considerations would apply to a non-magnetar
($B\lesssim10^{12}$G) neutron star with $\dot{m}=25$ since the spherisation radius (in cm) 
would be of the same order of magnitude as a $10M_{\odot}$ black hole as both $R_{\rm sph}$ 
and $T_{\rm sph}$ scale with the $\dot{M}$ and the mass of the compact object, whose trends 
nearly cancel out. This was briefly discussed in \citet{Pinto2020a}.}

The kinetic power of the wind can be written as
$L_w = 1/2 \, \dot{M}_w \, v_w^2 = 2 \, \pi \, m_p \, \mu \, \Omega \, C \, v_w^3 / \xi 
\, L_{\rm ion}  \, \sim 4 \times 10^{40} \, {\rm erg/s}$,
where $\dot{M}_w = 4 \, \pi \, R^2 \, \rho \, v_w^2 \, \Omega \, C$ is the outflow rate, $\Omega$ 
and $C$ are the solid angle and the volume filling factor (or \textit{clumpiness}), respectively,
which were adopted equal to 0.3 as conservative values from MHD simulations of winds driven 
by radiation pressure in super-Eddington winds \citep{Takeuchi2013},
$\rho$ is the density and $R$ is the distance from the ionising source.
Here we have used the $\xi$ definition to get rid of the $R^2 \rho$ factor
where $\rho=n_{\rm H} \, m_p \, \mu$ with $m_p$ the proton mass and $\mu = 0.6$ the 
average particle weight of a highly ionised plasma.

The filling factor of the wind might be much smaller. Using Eq. (23) in \citet{Kobayashi2018} 
and assuming that the outflow rate is comparable to the accretion rate, we obtain 
$C\sim3\times10^{-2}$. Systematics would tend to cancel out when also accounting for the
uncertainty on the ionisation parameter in the case for a harder SED ($\Delta$\,log\,$\xi \sim 1$).
In the pessimistic case the wind power would be of the order of $10^{39} \, {\rm erg/s}$,
which means still high enough to affect the surrounding medium and inflate ISM cavities.

The spectral shape, strong wind features, and presence of dips suggest that NGC 247 ULX-1 is 
likely being observed at high inclination (see Fig. \ref{Fig:big_picture}, left panel) where the funnel 
is already obscuring the innermost, hot, hard X-ray emitting regions.
As mentioned in Sect. \ref{sec:discussion_1keV}, the increase of the average flux level during the 
intermediate observations (3,4,5) might be caused by a higher local accretion rate.
Such a climate change would however affect the properties of both the disc and the wind.
The scale-height is already relevant around the Eddington limit (see, {e.g.}, \citealt{SS1973}, \citealt{Poutanen2007}). A further increase in the local $\dot{M}$ might push the optically-thick funnel
further upwards (see Fig. \ref{Fig:big_picture}, right panel) thereby obscuring the regions emitting 
photons with temperatures higher than that of the spherisation radius ($\gtrsim0.3$ keV), causing 
the very soft dips shown in Fig. \ref{Fig:Plot_XMM_lc} (see also \citealt{Urquhart2016}).

During the dips the high-ionisation portion of the wind could be hard to see as its absorption lines
were primarily affecting the (now) obscured hard X-ray continuum.
In fact, the strength of the high-ionisation (1.2-1.3 keV) absorption lines clearly decreases during 
the dipping observations (see Fig. \ref{Fig:Plot_EPIC_spectra_res}), while the 0.7 keV {\oviii} 
absorption line seems constant in flux if not even stronger. This might also suggest a stratification 
in the wind. Outside the dips, an overall increase in the accretion rate would also imply a stronger
radiative force and, therefore, a slightly faster wind which seems to be confirmed by the higher
blueshift of the lines (see Fig. \ref{Fig:Plot_EPIC_spectra} and \ref{Fig:Plot_EPIC_spectra_res}).
The 1 keV emission lines also weaken during the bright / dipping observations, 
indicating that they should be produced in the inner regions in agreement with their overall
larger broadening (see Table \ref{table:plasma_components}).

A similar picture was proposed by \citet{Guo2019} who argued that the $\sim$100s transitions 
can be explained by the viscous timescale with the X-ray flux variability driven by accretion rate
fluctuations (at $\dot{m}\gtrsim10$). However, local fluctuations in the $\dot{M}$ might also cause variations in the
winds, which could alter the source appearance \citep{Feng2016}.

Although fascinating and self-consistent, this scenario might be not the only one able
to explain all the observables.
Additional, alternative, and (ideally) model-independent approaches could be considered.
\textcolor{black}{For instance, another phenomenon which might explain the nature of the dips
might be the propeller effect due to a strong magnetic field. Such scenario would imply a decreasing
$\dot{M}$ and a geometrical beaming to cause the observed brightening.}
More insights on the temporal evolution of the spectral residuals accounting 
for the different spectral
states that the source shows inside / outside the dips will be given in D'A\`i et al. (in prep). Similarly,
the Fourier analysis of the characteristics timescales in NGC 247 ULX-1 and the corresponding 
association with the dipping activity will be shown by \citet{Alston2021}. 
Here, in particular, we argue that the alternation of the dips might be due to azimuthally-dependent structures.

We plan to investigate 
the variability of the RGS spectral lines to place more constraints onto their nature. However,
the low count rate of the grating spectra currently prevent us from trying to study them during the dips
and on timescales shorter than 100 ks in the bright states outside the dips.
Future missions like XRISM and Athena will revolutionise the study of ULX
thanks to their high effective area, high spectral resolution, and low background
(see, {\it e.g.}, \citealt{Barret2018}, \citealt{Guainazzi2018}).
\citet{Pinto2020b} simulated NGC 1313 ULX-1 microcalorimeter 
spectra for these two missions and showed 1) how XRISM will strengthen the 
identification of lines in the $1-10$ keV band and 2) how Athena / X-IFU will be able to detect
winds in observations with just 1 ks of exposure time. The latter is primarily due to the fact 
that X-IFU will have two orders of magnitude higher effective area than RGS.

\section{Conclusions}
\label{sec:conclusion}

Most ULXs are believed to be powered by super-Eddington accreting neutron stars 
and, perhaps, black holes. 
\textcolor{black}{The disc is expected to thicken at accretion rates above the Eddington rate}
and to launch powerful winds through radiation pressure and/or magnetic fields.
Evidence of winds has been found in several ULXs through high-resolution X-ray
spectrometers. 
It is yet unclear whether the switch between the classical soft and supersoft state -
which is observed in supersoft ULXs - is due to the thickening of the disc
and/or the optically-thick part of the wind.
In order to better understand such phenomenology and the overall super-Eddington 
mechanism, we undertook a large observing campaign with XMM-\textit{Newton} to study 
NGC 247 ULX-1, which is the brightest (in flux) of all supersoft ULXs.

The new observations showed for the first time unambiguous evidence of a wind in the form
of emission and absorption lines from highly-ionised ionic species, with the absorption phase
exhibiting a mildly-relativistic outflow ($-0.17c$) in line with the other ULXs whose grating spectra
had sufficient quality to detect and identify spectral lines. 
Remarkable variability was observed in the source flux with strong dipping activity during the 
brightest observations, which is typical among soft ULXs such as NGC 55 ULX-1, and indicate
a close relationship between the accretion rate and the appearance of the dips.
The latter are likely due to a thickening of the disc scale-height and the wind as shown by a 
progressively increasing blueshift in the spectral lines. 

%%%{\textcolor{red}{This was the first paper in a series of works focussing on our large 
%%%campaign on NGC 247 ULX-1 and forthcoming projects are being prepared with 
%%%primary focus on the broadband spectral evolution (D'A\`i et al. in prep) and the 
%%%properties of the timescales of the dips and the spectral continuum
%%%variability \citep{Alston2021}.}

\section*{Acknowledgments}

This work is based on observations obtained with XMM-\textit{Newton}, an
ESA science mission funded by ESA Member States and USA (NASA).
We acknowledge support from ESA Research Fellowships.
We thank the XMM-\textit{Newton} SOC for support in optimising our observing campaign
and J. M. Miller, D. Proga and M. Parker for useful discussion regarding winds 
and absorption in Galactic X-ray binary.
AD, MDS, EA  acknowledge financial support from the agreement 
ASI-INAF n.2017-14-H.0 and INAF main-stream.
\textcolor{black}{We thank the anonymous referee for their very useful suggestions.}

\section*{DATA AVAILABILITY}

All of the data and software used in this work are publicly available from ESA’s XMM-Newton Science Archive (XSA\footnote{https://www.cosmos.esa.int/web/xmm-newton/xsa)} and NASA’s HEASARC archive\footnote{https://heasarc.gsfc.nasa.gov/}. Our codes are publicly available and can be found on the GitHub\footnote{https://github.com/ciropinto1982}.

\bibliographystyle{mn2e}
\bibliography{bibliografia} %----> bibliografia.bib

\appendix

\section{Technical details}
\label{sec:appendix}

In this section we put technical detail, plots, and tables, that were excluded from the
main body of the paper. 

\subsection{Nearby bright X-ray source}
\label{sec:appendix_source_x2}

The RGS extraction region includes a few faint sources with the brightest one being 
XMMU J004710.0-204708 \textcolor{black}{(X-2 hereafter, see Fig. \ref{Fig:Plot_EPIC_image})}. 
We extracted \textcolor{black}{its} 
EPIC spectra from all observations and stacked them similarly to ULX-1.
The spectrum of X-2 is much flatter than that of the ULX-1 and can be well modelled with a powerlaw model ($\Gamma=1.60\pm0.03$), a moderate column density $N_{\rm H}=(1.0\pm0.1)\times10^{21}$\,cm$^{-2}$, and an intrinsic luminosity $L_{\rm 0.3-10 keV}=(1.45\pm0.04)\times10^{38}$\,erg/s (assuming a distance of 3.3 Mpc). This corresponds to the Eddington limit for a Solar-mass star and, given the spectral slope, the source X-2 could be a common XRB near the NGC 247 centre. At 1 keV its spectrum is remarkably featureless and 40-50 times fainter than ULX-1 implying that it will have no significant effects on the RGS spectral lines. 

\subsection{Modelling of individual EPIC observations}
\label{sec:appendix_observations}

\textcolor{black}{Table \ref{table:continuum_allabs} shows the results of the EPIC spectral modelling and the root-mean square estimated from the EPIC-pn data of each observation (see Sect. \ref{sec:residuals_variability} and \ref{sec:data_investigation}).}

\begin{table*}
\caption{Broadband properties of the individual observations.} %%%\textcolor{black}{Energies? Time-averaged SED?}}  
\label{table:continuum_allabs}     
   \vspace{-0.cm}
\renewcommand{\arraystretch}{1}
 \small\addtolength{\tabcolsep}{-1.6pt}
 
\scalebox{0.8}{%
\hskip-0.5cm\begin{tabular}{@{} l l l l l l l l l l l l l l}     
\hline  
\multicolumn{13}{c}{\multirow{1}{*}{\textcolor{black}{Spectra Model} : $hot\,(bb+bb) + (gaus+gaus+gaus)$}} & Lightcurve\\                                                                                                         
\hline
 & \multicolumn{2}{c}{\multirow{1}{*}{Blackbody 1}} & \multicolumn{2}{c}{\multirow{1}{*}{Blackbody 2}} & 0.3-10keV & \multicolumn{2}{c}{\multirow{1}{*}{Gaussian 1}} & \multicolumn{2}{c}{\multirow{1}{*}{Gaussian 2}} & \multicolumn{2}{c}{\multirow{1}{*}{Gaussian 3}}  & 0.3-10keV  & 0.3-10keV \\ 
\hline
Parameter     & Area              &   Temp              & Area              &   Temp          &  L$_{X,tot}$   &   Norm           & Energy          &   Norm           & Energy        &   Norm           & Energy         & $C/d.o.f.$ &  RMS        \\  
Units         & $10^{19}$cm$^2$ & $10^{-1}$\,keV   & $10^{16}$cm$^2$ & $10^{-1}$\,keV & $10^{39}$erg/s & $10^{46}$ph/s & $10^{-1}$\,keV & $10^{46}$ph/s & $10^{-1}$\,keV & $10^{45}$ph/s & $10^{-1}$\,keV&  &   (\%)       \\  
\hline
 0844860101 & $0.9\pm0.1$ & $1.38\pm0.02$ & $1.1\pm0.3$ & $3.7\pm0.1$ & $2.9\pm0.4$ & $-1.2\pm0.2$  & $6.9\pm0.1$ & $1.5\pm0.2$ & $9.5\pm0.1$ & $-6.9\pm0.8$ & $12.4\pm0.1$ & 321/160 & $16.1\pm 0.1$ \\
 0844860201 & $1.0\pm0.1$ & $1.31\pm0.02$ & $0.7\pm0.2$ & $3.5\pm0.2$ & $2.3\pm0.3$ & $-0.9\pm0.2$  & $6.7\pm0.1$ & $1.8\pm0.2$ & $9.3\pm0.1$ & $-5.1\pm0.6$ & $12.6\pm0.1$ & 341/140 & $11.7\pm 0.1$ \\
 0844860301 & $1.8\pm0.2$ & $1.21\pm0.02$ & $0.6\pm0.2$ & $3.3\pm0.2$ & $2.8\pm0.4$ & $-1.3\pm0.2$  & $6.6\pm0.1$ & $2.6\pm0.2$ & $9.1\pm0.1$ & $-6.6\pm0.5$ & $12.4\pm0.1$ & 441/128 & $  9.5\pm 0.1$ \\
 0844860401 & $0.6\pm0.1$ & $1.50\pm0.03$ & $1.0\pm0.1$ & $4.3\pm0.1$ & $2.7\pm0.3$ & $-1.4\pm0.2$  & $7.5\pm0.1$ & $0.8\pm0.2$ & $9.9\pm0.1$ & $-2.8\pm1.0$ & $12.1\pm0.2$ & 412/192 & $47.7\pm 0.2$ \\
 0844860501 & $0.6\pm0.1$ & $1.53\pm0.03$ & $1.0\pm0.2$ & $4.2\pm0.1$ & $2.9\pm0.3$ & $-1.7\pm0.2$  & $7.4\pm0.1$ & $0.6\pm0.2$ & $9.8\pm0.2$ & $-3.0\pm0.9$ & $13.0\pm0.2$ & 391/188 & $26.9\pm 0.2$ \\
 0844860601 & $0.5\pm0.1$ & $1.47\pm0.03$ & $0.6\pm0.2$ & $4.1\pm0.2$ & $2.2\pm0.3$ & $-1.1\pm0.2$  & $7.2\pm0.2$ & $0.6\pm0.2$ & $9.8\pm0.2$ & $-3.7\pm0.9$ & $12.6\pm0.2$ & 271/154 & $54.7\pm 0.2$ \\
 0844860701 & $1.8\pm0.2$ & $1.20\pm0.02$ & $0.2\pm0.1$ & $3.9\pm0.3$ & $2.8\pm0.3$ & $-1.1\pm0.3$  & $6.5\pm0.2$ & $2.2\pm0.2$ & $9.0\pm0.1$ & $-5.8\pm0.6$ & $11.8\pm0.2$ & 324/122 & $11.3\pm 0.1$ \\
 0844860801 & $1.8\pm0.2$ & $1.21\pm0.02$ & $0.5\pm0.3$ & $3.4\pm0.3$ & $2.8\pm0.4$ & $-1.9\pm0.3$  & $6.6\pm0.1$ & $2.0\pm0.2$ & $9.2\pm0.1$ & $-7.2\pm0.7$ & $12.2\pm0.1$ & 298/109 & $11.3\pm 0.1$ \\
%$L_{X\,mbb1}$                                    &  $3.1 \pm 0.4$      &  $2.4 \pm 0.3$      &  $5.7 \pm 0.4$       \\               
%$L_{X\,mbb2}$                                    &  $3.5 \pm 0.5$      &  $3.5 \pm 0.7$      &  $1.5 \pm _{1.5}^{2.5}$       \\  
%$L_{X\,comt}$                                     &  $3.7 \pm 0.8$      &  $3.6 \pm 0.9$      &  $7.9 \pm 0.9$     \\  
%
%$kT_{mbb1}$ (keV)                              &  $0.37 \pm 0.02$  &  $0.37 \pm 0.02$  &  $0.5 \pm 0.1$   \\              
%$kT_{mbb2}$ (keV)                              &  $1.5  \pm  0.2  $  &  $1.6  \pm  0.2  $  &  $1.3  \pm  0.5  $   \\     
%$kT_{in,\,comt}$\,(keV)                        &  $1.5^{(c)}$           &  $1.6^{(c)}$            &  $1.3^{(c)}$            \\                
%$kT_{e,\,comt}$\,(keV)                         &  $3.8  \pm  0.2 $   &  $3.9  \pm  0.2 $   &  $4.1  \pm  0.1 $    \\                
%$\tau_{comt}$                                      &  $5.1  \pm 0.4 $    &  $5.1  \pm 0.5 $    &  $3.9  \pm 0.1 $   \\                
% 
%$N_{\rm H}\,(10^{21} {\rm cm}^{-2})$  &   $1.9^{(c)}$           &  $1.9 \pm 0.1$      &     $1.9^{(c)}$        \\  
%$C$-stat/d.o.f.                                     &    779/614               &    789/604              &    809/614              \\               
\hline                                                                                                                
\end{tabular}}

$L_{X\,(0.3-10\,\rm keV)}$ luminosities are calculated assuming a 
distance of 3.3\,Mpc and are corrected for absorption (or de-absorbed,
%%%The grouped spectra are defined in Sect.\,\ref{sec:fluxed_spectra}.
see Fig.\,\ref{Fig:Plot_EPIC_spectra}).
%%%while $N_{\rm H}$ is coupled between the observations.
   \vspace{-0.2cm}
\end{table*}

\subsection{\textcolor{black}{CIE model scan for the} SS 433 RGS spectrum}
\label{sec:appendix_ss433}

\begin{figure}
\centering
  \includegraphics[width=0.9\columnwidth, height=5cm, angle=0]{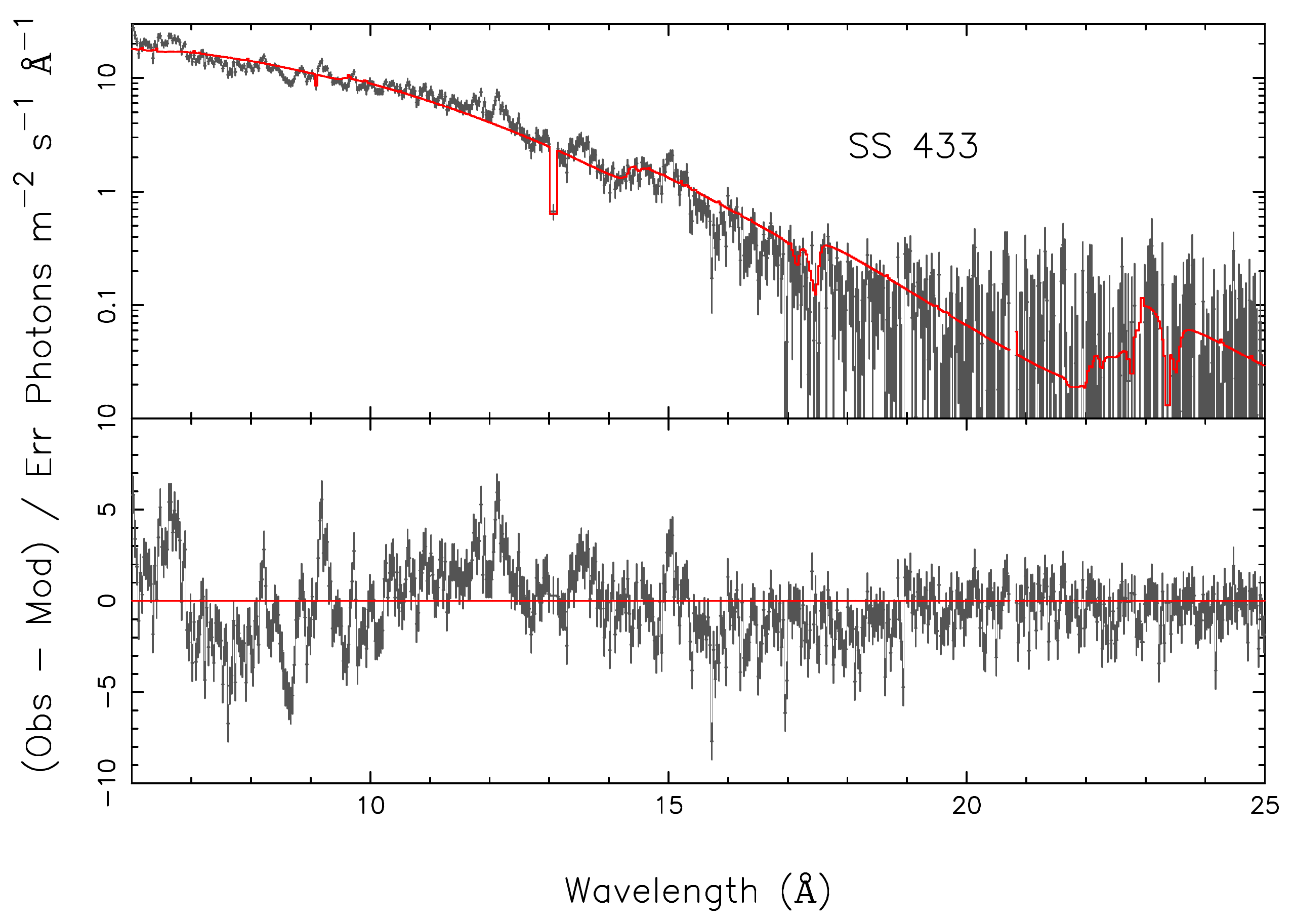}
  \vspace{-0.3cm}
   \caption{SS\,433 RGS spectrum and continuum model.}
   \label{Fig:ss433_continuum}
  \vspace{-0.3cm}
\end{figure}

We analysed the XMM-\textit{Newton} RGS spectrum of SS 433 
from the observation id:0694870201 (2012-10-03),
which provides the longest ($\sim$130\,ks) 
and best-exposed RGS spectrum of SS 433.
We reduced the RGS spectrum of SS 433 obsid 0694870201
identically to that of the NGC 247 ULX-1 data shown in Sect.\,\ref{Sect:rgs_reduction}. 
After removing the very little Solar flares we are left with 129.4\,ks for both RGS 1 and 2
cameras. We used the 6--25\,{\AA} range because at higher wavelengths 
the source emission is significantly absorbed and the background noise dominates 
the RGS spectrum (see Fig.\,\ref{Fig:ss433_continuum}).
No significant pile up was found in the RGS spectra.

Strong emission lines were observed close to the rest-frame energies of the 
most relevant transitions such as {\sixiii}, {\mgxi}, {\nex}, {\neix}, {\fexvii}, and {\oviii},
which is very similar to NGC247 ULX-1, albeit at higher significance because 
SS 433 is much closer ($\sim5$ kpc) and brigther.
We modelled the RGS spectral continuum with an absorbed powerlaw model
obtaining results similar to \citet{Marshall2013} and \citet{Medvedev2018}
such as a column density $N_{\rm H} = (1.14\pm0.02)\times10^{22}{\rm cm}^{-2}$, 
a slope $\Gamma=2.53\pm0.06$ and an X-ray unabsorbed luminosity 
$L_{\rm [0.3-10\,keV]} = (1.03\pm0.05)\times10^{36}{\rm erg/s}$.
We obtained high C-stat/d.o.f = 2393/618 as expected, due to the strong,
unmodelled, emission lines.

\begin{figure}
\centering
\includegraphics[width=0.95\columnwidth, angle=0]{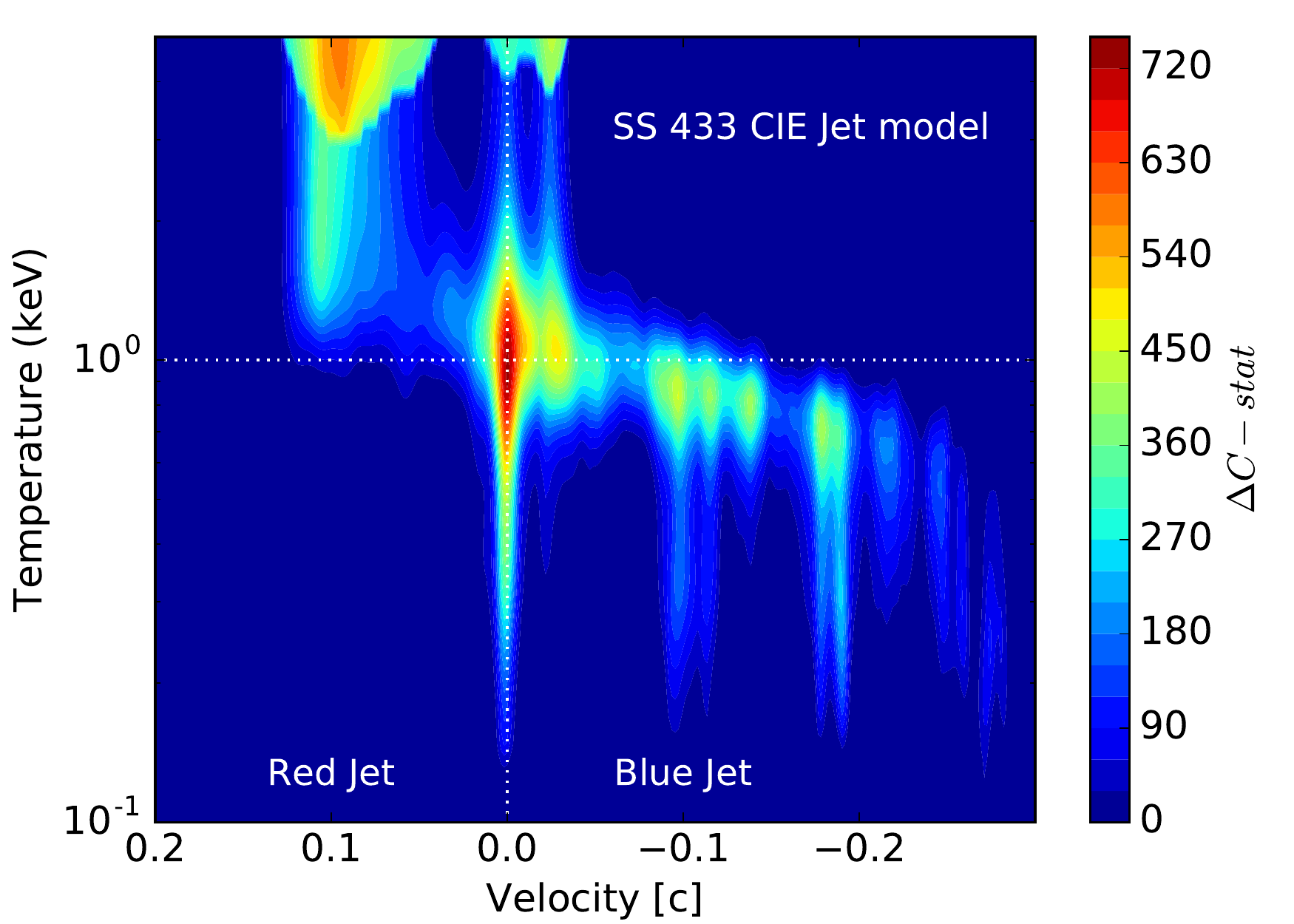}
   \vspace{-0.3cm}
  \caption{Multi-dimensional scan of collisional-ionisation emission model for 
                the SS 433 RGS spectrum.
                The X-axis shows the line-of-sight velocity.
                \textcolor{black}{Labels are same as in Fig. \ref{Fig:emitting_gas_CIE}}.}
   \label{Fig:ss433_cie_scan}
   \vspace{-0.3cm}
\end{figure}

\textcolor{black}{We tested onto the SS 433 RGS spectrum the same routine 
used for the NGC 247 ULX-1 data in Sect. \ref{sec:appendix_hot} to check the robustness of our method. 
We adopted collisional ionisation equilibrium ($cie$ model in {\scriptsize{SPEX}}) to model the jet emission.}
The velocity dispersion was fixed to 500 km/s, {\it i.e.} close to the RGS spectral resolution, 
the abundances were chosen to be Solar (to limit the computing time). We found a dominant 
low-velocity solution with an average temperature of 1 keV (see dotted horizontal and vertical lines
in Fig.\,\ref{Fig:ss433_cie_scan}).
The velocity is consistent with the dynamical range found by \citet{Medvedev2018} using the Fe K 
lines from the EPIC-pn spectrum, but showed a lower temperature, which is expected given 
that the RGS spectrum is more sensitive to the cooler gas phase of the multi-temperature jet. 

\subsection{SED modelling and \textcolor{black}{systematics effects}}
\label{sec:appendix_SED}

The non detection of the optical and UV counterpart of NGC 247 ULX-1 (Sect. \ref{sec:om_reduction})
prevent us from building a simultaneous multi-wavelength SED for our source.
This might have systematic effects on the calculation of the photoionisation balance.
Moreover, the flux upper limits obtained with the OM suggest that at least in the far-UV energy band
the source flux was lower than the levels measured in the archival HST observations that we used here.

\begin{figure}
 \centering
  \includegraphics[width=0.87\columnwidth, height=4.9 cm, angle=0]{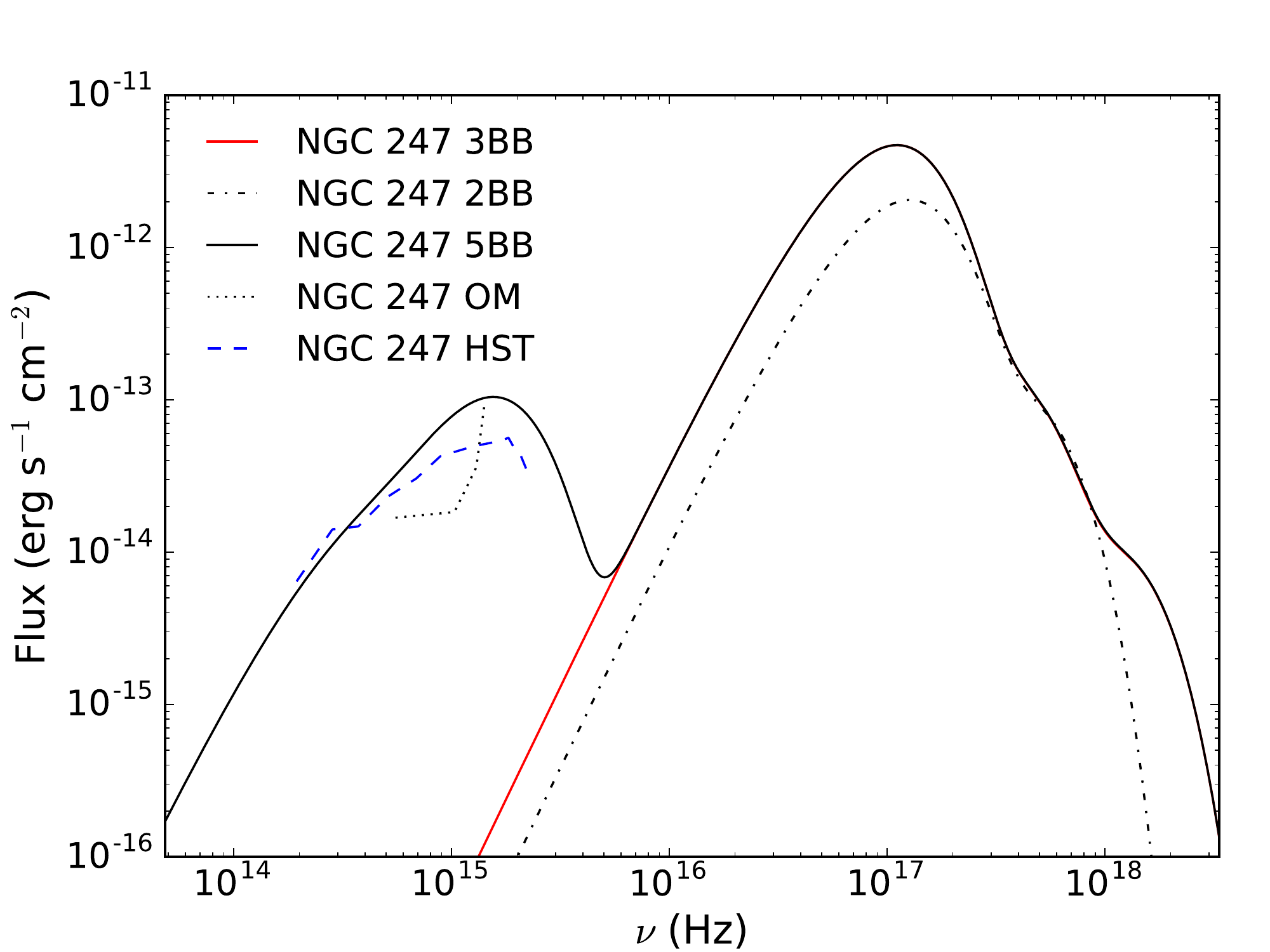}
  \vspace{-0.1cm}
  \includegraphics[width=0.91\columnwidth, height=4.3 cm, angle=0]{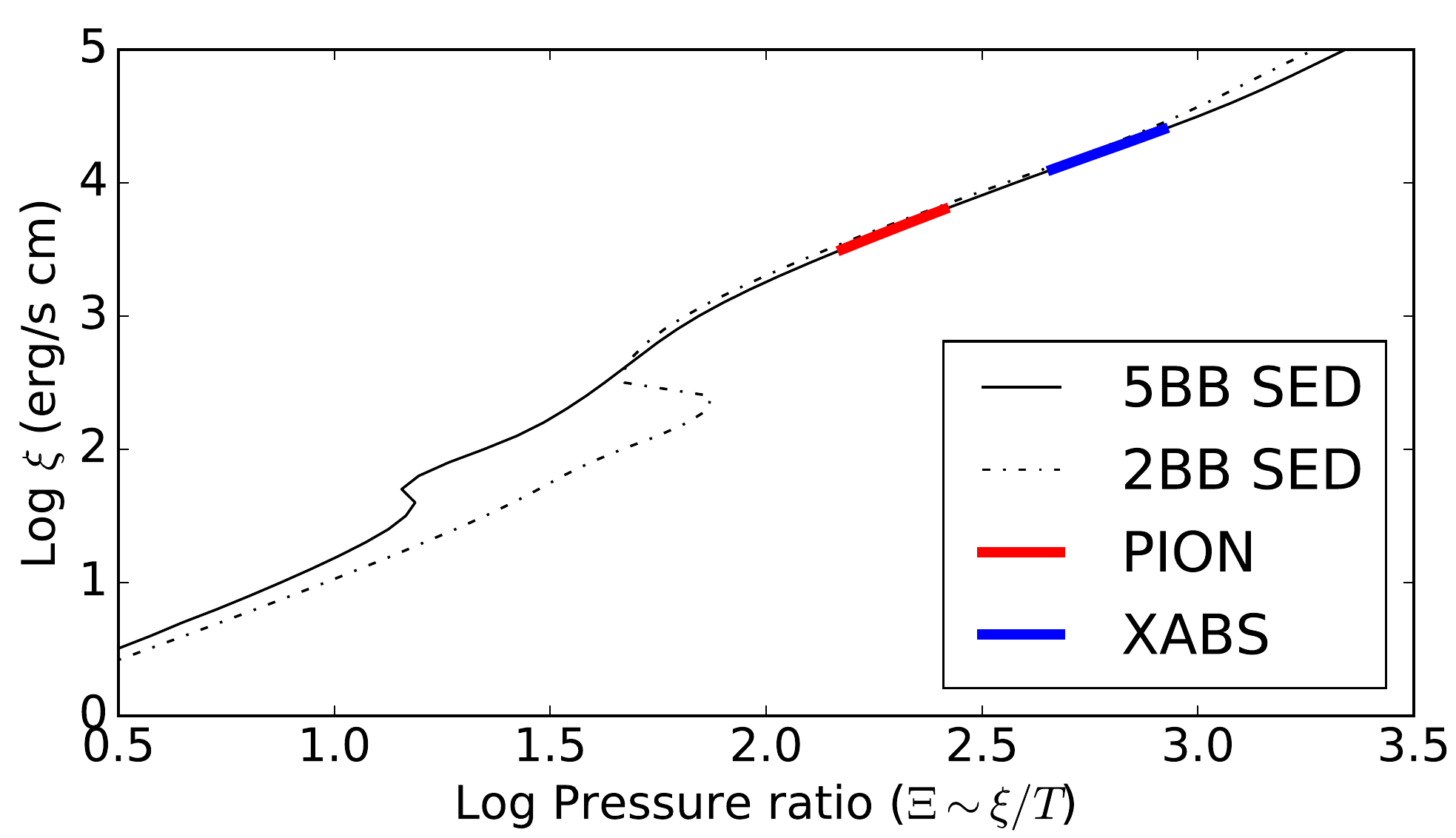}
  \vspace{-0.1cm}
   \caption{\textcolor{black}{SED (top panel)
                 and thermal-stability curve (bottom panel) computed for the time-averaged spectrum 
                 of NGC 247 ULX-1. The baseline SED (solid black line) consists of five unabsorbed 
                 blackbody models that account for emission in the optical, UV and X-ray bands. 
                 An alternative, simpler SED model uses just the X-ray two-blackbody model
                 (dashed-dotted line).}}
   \label{Fig:Plot_SED_Balance}
  \vspace{-0.4cm}
\end{figure}

\textcolor{black}{\citet{Pinto2020a} showed that a lower IR-to-UV flux in moderately-hard sources,
such as NGC 1313 ULX-1, mainly strengthens thermal instabilities at intermediate temperatures 
and ionisation parameters.
In Fig. \ref{Fig:Plot_SED_Balance} we show the SED adopted here (solid black curve)
consisting of a 5-blackbody model along with the simple 2-blackbody model
(dashed-dotted black line, see also Sect. \ref{sec:SED}).
The lower panel shows the stability curves
computed for these models.} Some deviations are mainly seen
at log $\xi$ from $1.5-2.5$, which is well below the values measured in this work
(see Table \ref{table:plasma_components}). This is not surprising given the shortage
of hard X-ray in our spectra which are the primary responsible of thermal instabilities.
This suggests that the ionisation balance is not significantly affected even if the
optical / UV fluxes were 2 orders of magnitude lower than our assumptions.

\subsection{Monte Carlo simulations and significance}
\label{sec:mc_simulations}

The $\Delta C_{\rm max}$ improvement to the continuum model does not necessarily yield 
the significance of the corresponding emission or absorption line models. This is 
due to the large parameters space that was explored and the possibility of 
detecting random spectral features (the look-elsewhere effect).

Among our physical model searches, the one for the absorption lines provided the 
smallest $\Delta C_{\rm max}$ due to their strength being lower than that typical
of the emission lines. We therefore focused on the 
results obtained with the \textit{xabs} component and used them as a proxy for the \textit{pion}.

\begin{figure}
 \centering
  \includegraphics[width=0.95\columnwidth, angle=0]{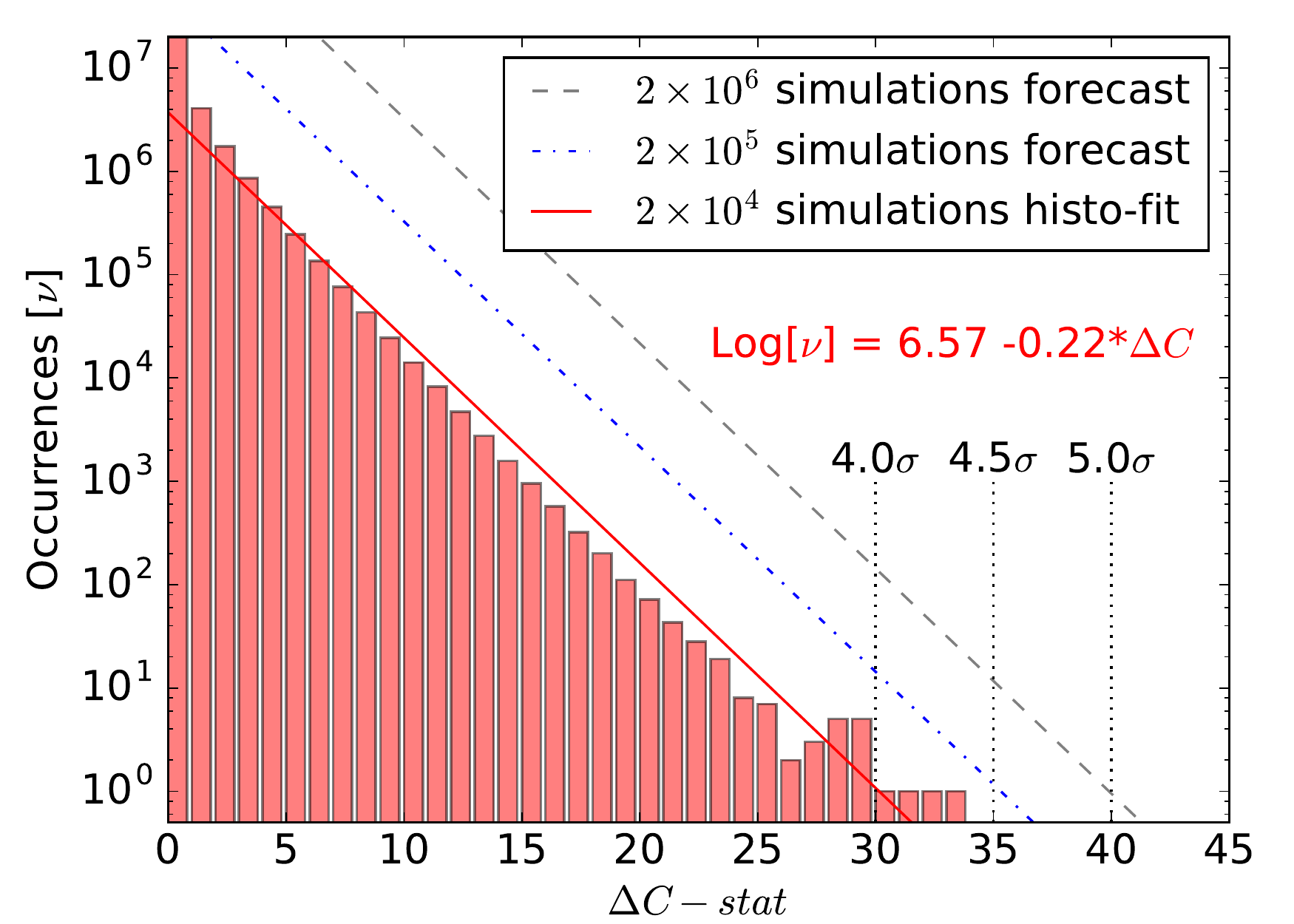}
  \vspace{-0.1cm}
   \caption{\textcolor{black}{Histogram and corresponding power-law fit of the 20k Monte Carlo simulations
                 of NGC 247 ULX-1 and forecast for 200\,000 and 2 million simulations.}}
   \label{Fig:MC_simulations}
  \vspace{-0.4cm}
\end{figure}

Following the method used in \citet{Pinto2020b}, we simulated 20\,000 RGS and EPIC
spectra adopting the 3-blackbody continuum model. Each faked spectrum was
scanned with the same \textit{xabs} grids used in Sect.\,\ref{sec:xabs_gas}.
The results of our MC simulations are shown in Fig.\,\ref{Fig:MC_simulations}. 
No outlier was found with $\Delta C \geq \Delta C_{\rm max} = 46$,
which suggests a significance $>4\sigma$ for the absorbing gas.

\textcolor{black}{We compared the simulations histogram for NGC 247 ULX-1 with those obtained 
for different sources by adopting a similar approach:}
20k simulations of NGC 1313 ULX-1 \citep{Pinto2020b}, 
2k for NGC 5204 ULX-1 \citep{Kosec2018a}
and 50k for the same data with a new, faster, cross-correlation method (Kosec et al.
submitted), 20k for ULX NGC 7793 P13 (Pinto et al. in prep), and 1k for AGN PG 1448
\citep{Kosec2020b}.
%%%Each run is computationally expensive, but very accurate in estimating the significance.
%%%The parameter search spaces have different sizes among the different 
%%%works due to the particular choice of model, parameter type and step
%%%({\it e.g.}, \% of energy points, \% of log $xi$ values, \% of $kT$ values, \% of simulations, etc.).
%%%More detail on these runs is provided in Appendix \ref{sec:appendix_MC}.
We fit the histograms of the logarithm of the occurrences
with straight lines and found an average slope $\overline{\Gamma}=-0.225\pm0.015$, 
which agrees with the simulations of NGC 247 ULX-1 
($\Gamma=-0.218\pm0.006$).

\textcolor{black}{We used the best-fit straight lines} to estimate the overall 
shape of the $\Delta C$-stat distribution
and forecast the results of larger numbers of simulations 
thanks to the agreement between the trends from 1\,000 to 50\,000 simulations. 
We therefore scaled the histogram fit of NGC 247 ULX-1,
assuming a constant slope and
multiplying the intercept of the straight line by a number equal to the ratio of the parameter
space that we want to forecast for a given number of simulations and the one we
obtained with 20\,000 simulations. In Fig. \ref{Fig:MC_simulations} we show the
predictions for $2\times10^5$ (dash-dotted line) and  $2\times10^6$ (dashed line) simulations.
This would suggest 4.5 and 5\,$\sigma$ detection probabilities for $\Delta C$-stat above
35 and 40, respectively, in the data with an uncertainty of $0.2\sigma$ according to the spread  
in the slope of the other histograms.

We finally retrieved the various $\Delta C$-stat values that correspond to
confidence levels ranging from 2.0, 2.5, ..., 5.0\,$\sigma$ and plot them as black contours
in Fig. \ref{Fig:photoionised_gas}. The $\sigma$ contours for the {\it pion} model scan were 
calculated by scaling the parameter space in the histogram of the {\it xabs} simulations
in the same way used for the forecast. 
%%%The main difference between {\it pion} and {\it xabs}
%%%model scan is that the former has a twice larger parameter search space due to the redshift
%%%and blueshift velocity interval, which was account for.

%%%\bsp

\label{lastpage}

\end{document}